\documentclass[nofootinbib,showpacs,superscriptaddress,titlepage,floatfix,amsmath,amssymb,preprint]{revtex4-1}
\usepackage{graphicx}
\usepackage{dcolumn}
\usepackage{subfigure}
\usepackage{bm}

\usepackage{amsmath} 
\usepackage{comment}
\newcommand{\bmath}{\begin{displaymath}}
\newcommand{\emath}{\end{displaymath}}
\usepackage[T1]{fontenc}
\DeclareMathOperator\arctanh{arctanh}

\def\Xint#1{\mathchoice
   {\XXint\displaystyle\textstyle{#1}}%
   {\XXint\textstyle\scriptstyle{#1}}%
   {\XXint\scriptstyle\scriptscriptstyle{#1}}%
   {\XXint\scriptscriptstyle\scriptscriptstyle{#1}}%
   \!\int}
\def\XXint#1#2#3{{\setbox0=\hbox{$#1{#2#3}{\int}$}
     \vcenter{\hbox{$#2#3$}}\kern-.5\wd0}}

\def\dashint{\Xint-}

\newcommand{\la}{\label}
\newcommand{\be}{\begin{equation}}
\newcommand{\ee}{\end{equation}}
\newcommand{\bea}{\begin{eqnarray}}
\newcommand{\eea}{\end{eqnarray}}

\newcommand{\p}{\partial}
\newcommand{\ba}{\begin{align}}
\newcommand{\ea}{\end{align}}

\usepackage{color}

\def\Xint#1{\mathchoice
   {\XXint\displaystyle\textstyle{#1}}%
   {\XXint\textstyle\scriptstyle{#1}}%
   {\XXint\scriptstyle\scriptscriptstyle{#1}}%
   {\XXint\scriptscriptstyle\scriptscriptstyle{#1}}%
   \!\int}
\def\XXint#1#2#3{{\setbox0=\hbox{$#1{#2#3}{\int}$}
     \vcenter{\hbox{$#2#3$}}\kern-.5\wd0}}

\def\dashint{\Xint-}

\begin{document}

\title{Hydrodynamics of local excitations after an interaction quench 
in $1D$ cold atomic gases}

\author{Fabio Franchini}
\email{gfabio@gmail.com}
\affiliation{Ru{\dj}er Bo\v{s}kovi\'c Institute, Bijen\v{c}ka cesta 54, 10000 Zagreb, Croatia}
\affiliation{INFN, Sezione di Firenze, Via G. Sansone 1, 50019 Sesto Fiorentino (FI), Italy}
\affiliation{The Adbus Salam ICTP, Strada Costiera 11, 34151, Trieste, Italy}

\author{Manas Kulkarni}
\email{mkulkarni@citytech.cuny.edu}
\affiliation{Department of Physics, New York City College of 
Technology, The City University of New York, Brooklyn, NY 11201, USA}

\author{Andrea Trombettoni}
\email{andreatr@sissa.it}
\affiliation{CNR-IOM DEMOCRITOS Simulation Center, Via Bonomea 265, 
I-34136 Trieste, Italy}
\affiliation{SISSA and INFN, Sezione di Trieste, Via Bonomea 265, 34136, Trieste, Italy}

\begin{abstract}
We discuss the hydrodynamic approach to the study of the time evolution 
-- induced by a quench -- of local excitations in one dimension. 
We focus on interaction quenches: the considered protocol consists in 
creating a stable localized excitation propagating through the system, and then operating 
a sudden change of the interaction between the particles. To 
highlight the effect of the quench, 
we take the initial excitation to be a soliton. The quench splits the excitation into two packets 
moving in opposite directions, whose characteristics can be expressed in a universal way. 
Our treatment allows to describe the internal dynamics of these two packets 
in terms of the different 
velocities of their components. We confirm our analytical predictions 
through numerical simulations 
performed with the Gross-Pitaevskii equation 
and with the Calogero model (as an example of 
long range interactions and solvable with a parabolic confinement). 
Through the Calogero model we also discuss 
the effect of an external trapping on the protocol. 
The hydrodynamic approach shows that there is a difference between 
the bulk velocities of the 
propagating packets and the velocities of their peaks: 
it is possible to discriminate the two quantities, as we show through the comparison between numerical simulations and analytical estimates. In the realizations of the discussed quench protocol in a cold atom experiment, these different velocities are accessible through different measurement procedures.
\end{abstract}

\pacs{03.75.Kk, 03.75.Lm, 05.45.Yv, 67.85.-d, 05.70.Ln, 02.30.Ik}

\date{\today}

\maketitle

\tableofcontents

\section{Introduction}

The outstanding performances of modern experiments in preparing and controlling setups of quantum gases \cite{bloch2008,lamacraft2011} 
have promoted the study of out-of-equilibrium systems in ultracold gases 
as one of the arguably most challenging topics in the field  
\cite{greiner2002,kinoshita2006,joerdens2008,chen2011,trotzky2012,gring2012,cheneau2012,schneider2012,langen2013,smith2013,pagano2014,braun2015,vidmar2015,preiss2015,steffens2015,langen2015}. 
One of the reasons for the complexity of this challenge is due to 
the variety of ways in which a system can be driven out of equilibrium and 
the difficulty in having a general guidance principle to relate their phenomenologies. 

Thus, over the years, the community has concentrated on a few protocols that have emerged to be sufficiently clean and interesting. One question at the forefront has been under which conditions (and in which sense) a system is able to reach an equilibrium, and whether this equilibrium can show universal characteristics such as thermalization 
\cite{jaynes1957,deutsch1991,srednicki1994,calabrese2007a,rigol2007,gambassi2012,essler2012,pozsgay2014,ilievski2015,goldstein2015,gogolin2015}. Particularly useful in addressing such questions are setups based on cold atoms 
\cite{degrandi2010,camposvenuti2010,iyer2012,karrasch2012,dallatorre2013,franchini2015,piroli2016}.

One of the experimentally most relevant protocol to study out-of-equilibrium dynamics and the issue of thermalization is the 
quench protocol, in which, typically, a system described by a Hamiltonian $H$ 
is prepared in its ground state and then, at a given moment of time, 
let evolved using a different Hamiltonian $H^\prime$ \cite{polkovnikov2011}.  
The questions usually asked with quantum quenches refer to late times properties of the systems (and on whether the unitary evolution can lead to something that can locally be described in a semi-classical way) 
\cite{rigol2009,biroli2010,canovi2012,calabrese2012a,wouters2014}.

Variations to the usual quench protocol have been considered mainly 
in two forms: the initial state was taken to be an excited state 
of the initial Hamiltonian \cite{bucciantini2014}, or the interaction was changed only locally 
\cite{silva2008,diez2010,stephan2011,munder2011,torres2014}. 
In a recent work \cite{franchini2015}, we introduced a different protocol: 
we proposed to start with a local excitation and to let it evolve after 
a global interaction quench. In order to isolate the effect of the 
quench on the dynamics, the initial state is prepared as a solitonic 
excitation of the original Hamiltonian, that is, a state for 
which the one-particle density profile evolves without (almost) changing its shape. This setup has the merit of being experimentally feasible and of showing universal properties already at short times. In \cite{franchini2015} it was also shown that, by changing the underlying interaction during the soliton motion, 
the excitation breaks into two profiles: one moving in the same direction 
as the initial excitation, the other in the opposite. 
In a cold atom setup, such a quench can be triggered through the trapping 
or with an external magnetic field \cite{bloch2008} to induce a change 
in the scattering length and in the speed of sound.
The system is then let evolve for short times after the quench and the 
velocities and shapes of the two chiral profiles created by the quench 
can be measured either by direct imaging or reconstructed 
by releasing the trap, through time-of-flight measurements.
 
As we are going to discuss in this paper, a convenient way to study 
the dynamics of a local excitation 
is to employ a (non-linear) hydrodynamics description of the system 
in terms of the density and velocity fields. Moreover, 
if the initial density profile is not a big perturbation over the background, 
velocities and shapes of the transmitted and reflected profiles 
for short times after the quench can be expressed in a universal way, 
that is independent from the details of the quench and of the microscopic 
interaction \cite{franchini2015}. The hydrodynamic approach \cite{forster} 
is clearly complementary to a microscopic computation 
of the dynamics (when practically doable), and it has the advantage that microscopic details of the 
underlying model enter as parameters of the hydrodynamic equation (e.g., 
the sound velocity). It also provides a standard tool to study 
collective excitations and dynamical properties in cold atom setups 
\cite{stringari,kulkarni2012} and it applies as well to higher dimension: 
however 
in this paper, in line with the topic of the Focus Issue and with 
our choice of consider solitonic solutions
we limit ourself to one dimensional systems and soliton excitations (in higher dimensions, solitonic states are stable only for limited times and one needs to take into account the spreading of the wave packet). 
We consider in detail the Gross-Pitaevskii (GP) equation, 
with $\delta$-like interactions (which is relevant for the 
$1D$ Bose gases in the limit of small interactions \cite{stringari}), 
and the Calogero model, since it is 
exactly solvable also with in the presence of a parabolic trapping potential  
\cite{abanov2011,kulNPB,rajabpour2014}. 

The same quench setup has also been considered in \cite{gamayun2014,gamayun2015,caudrelier2015} with a specific interest on integrable PDEs 
describing the evolution of one dimensional systems and on the translation of the quench protocol 
in the corresponding quantum inverse scattering problem. 
These papers signal the interest of the community in the interaction 
quench. We would like to stress that the proposed quench protocol 
could be implemented relatively easily in cold atoms experiments 
and that, by focusing on the short time universal dynamics, 
our predictions can be tested directly in the laboratory, 
which should be contrasted with typical large time results 
produced by other approaches and protocols. 
Universal properties of short time out-of-equilibrium dynamics 
has been considered  
in \cite{franchini2015,morawetz2014,chiocchetta2015,chiocchetta2016}.

In this work our main goal is to investigate in detail the hydrodynamic 
approach for the study of the dynamics of solitonic excitations in 
one-dimensional systems, clarifying the hypothesis behind the derivations based on 
the hydrodynamic approach and to discuss on the possible experimental 
realization of the quench protocol in cold atom systems. To the latter 
aim, we address an important 
issue related to the measurability of the chiral profiles generated by the quench and relevant for the cold atom physical realizations. 
In fact, at the quench time the transmitted and reflected profiles 
are perfectly overlapping and can be distinguished only after they have 
moved apart. While in a time-of-flight experiments one need not wait 
this time, since the two profiles have opposite momentum, in 
a real imaging scenario (such as that we employ in our numerical experiments) 
this waiting time can introduce additional effects. We analytically estimate this waiting time.
Moreover, since the two chiral profiles are not solitons of the post-quench 
system, the center of mass (average) velocity of each profile might 
be different from the velocity of its highest point. We find that in the 
hydrodynamic approach one can naturally introduce both the 
``bulk'' and ``peak'' velocities of the transmitted and reflected packets: 
we provide expressions for both quantities, and we compare them with 
our numerical simulations for the GP and Calogero models. We also discuss the time scales relevant for the experimental realization of the proposed protocol and the efficacy of our approximation within them.

The plan of the paper is the following. In Section \ref{sec:hydro} 
we review the hydrodynamic approach to the description of one dimensional 
cold atomic gases \cite{kulkarni2012}. In Section \ref{sec:quench} 
we analyze the quench protocol and derive the expressions 
for the velocities and heights of the chiral profiles generated by the quench. 
In Sections \ref{sec:NLSquench} and \ref{sec:Calogeroquench} 
we compare the analytical prediction against our numerical simulation performed 
on the one-dimensional GP equation and on the Calogero model. 
In Section \ref{sec:conclusions} we discuss our results and comment 
on future perspectives. Finally, we collect some useful 
(although more technical) information in the Appendices. 
Appendix \ref{app:KdVreduction} explains how we can extract 
an integrable dynamics out of a generic hydrodynamic system, while 
in Appendix \ref{app:KdV} we discuss the main properties of this 
integrability, known as the Korteweg-de Vries (KdV) equation. 
In Appendices \ref{app:NLSE} we discuss 
some different versions of the $1D$ GP and their relation to KdV. 
Finally, in Appendices \ref{app:Calogero} and \ref{app:HarmCal} 
we collect the basic facts about the rational Calogero model and its 
version confined in a harmonic potential (also integrable).

\section{The hydrodynamic approach}
\label{sec:hydro}

All quantum systems at sufficiently low temperature acquire a collective 
behavior. In many such cases, the expectation value of the particle 
density operator $\hat{\rho} (x) = \sum_j \delta(x-x_j)$ 
becomes a smooth function and the quantum fluctuations around it are 
negligible. When this happens, the system dynamics can be described 
with a hydrodynamic approach \cite{forster}. 
In the following, we will primarily be interested 
in one dimensional systems with one particle species 
without additional internal degrees of freedom, 
so that it is sufficient to introduce the scalar density 
$\rho(x,t)$ and velocity $v(x,t)$ fields.

Assuming Galilean invariance and locality, the most general hydrodynamic Hamiltonian reads \cite{kulkarni2012}
\be
   H = \int d x \left[ { \rho v^2 \over 2 } + \rho \epsilon(\rho) 
   + A(\rho) {\left( \partial_x \rho \right)^2 \over 4 \rho} \right] \; ,
\ee
where $\epsilon(\rho)$ is the internal energy and $A(\rho)$ is related to quantum pressure. We also assume that dissipative effects can be neglected. The equations of motions can be obtained by remembering that the density and velocity are conjugated fields \cite{landau,khalatnikov} satisfying 
\be
  \Big\{ \rho(x), v(y) \Big\} = \partial_x \delta (x-y) \;.
\ee
They are the continuity and Euler equations \cite{forster}:
\bea
\label{cont}
&& \dot \rho + \p (\rho v) = 0; \: \: \: \: \dot v + \p {\cal A}=0\quad \\
\label{Euler} 
&& {\cal A} \equiv \frac{v^2}{2} + \omega(\rho) - A^\prime(\rho)(\p\sqrt{\rho})^2 - A(\rho)       
\frac{\p^2\sqrt{\rho}}{\sqrt{\rho}},
\eea
where $\omega (\rho) = \partial_\rho \big( \rho \epsilon(\rho) \big)$ is the specific enthalpy (which at zero temperature reduces to the chemical potential).

Under the foretold conditions, these equations are completely general: the functions $\epsilon(\rho)$ and $A(\rho)$ encode the specificity of the system under consideration. A discussion of how one can extract an integrable dynamics 
out of a generic hydrodynamic system and the derivation of the KdV 
(used in the next Section) is reported in Appendix 
\ref{app:KdVreduction}, and useful 
facts about the KdV are in Appendix \ref{app:KdV}.

Choosing 
\be
   \epsilon = {g \over 2 m} \rho \; , \qquad
   A = {\hbar^2 \over 2 m} \:,
   \label{NLSEparameters}
\ee
and combining the hydrodynamic fields into the complex field
\be
   \Psi = \sqrt{\rho} e^{i {m \over \hbar} \int^x v(y) d y}
   \label{psiansatz}
\ee
the dynamical equations (\ref{cont},\ref{Euler}) can be written as the single complex equation
\be
i \hbar \partial_t \Psi (x,t) = \left\{ - {\hbar^2 \over 2 m} \partial_{xx} 
+ g \left( \left| \psi (x,t) \right|^2 - \rho_0 \right) \right\} \psi(x,t) \; ,
\label{NLS}
\ee 
which can be recognized as the $1D$ non-linear Schr\"odinger equation, 
aka the $1D$ GP equation [without the external potential 
in (\ref{NLS})]. In Eq. (\ref{NLS}) $\rho_0$ is the density for $x\to\infty$.
Details about the the non-linear Schr\"odinger equation and its KdV reduction 
are in Appendix \ref{app:NLSE}.

The GP equation has been routinely used in the last two decades 
to describe the dynamics of ultracold bosons at $T=0$ \cite{stringari}. 
In $3D$ its validity stems from the fact that there is a condensate, whose 
macroscopic wavefunction obeys the GP equation, from which one can derive 
hydrodynamic equations \cite{stringari}. In $1D$ there is no condensate, 
since there is actually a ``quasi-condensate'' \cite{castin2004}: however, 
the GP gives a good description of experimental results 
for small interactions (i.e., for the coupling constant of the Lieb-Liniger 
model, defined in (\ref{gammadef}), $\gamma =\frac{m}{\hbar^2} \frac{g}{\rho_0} \ll 1$ \cite{pethick}) and in particular of the 
-- dark and bright -- soliton dynamics observed in $1D$ experiments 
\cite{burger1999,khaykovich2002} and of shock waves 
\cite{meppelink2009}. When in $1D$ the coupling constant 
is not much smaller than $1$, one can nevertheless use 
hydrodynamic classical equations (as the ones given above) 
to study small deviations from equilibrium \cite{menotti2002}. For larger 
deviations one has to study the quantum dynamics 
using directly the Lieb-Liniger model 
(which might be at present rather challenging 
from the computational point of view) or resort to 
$1D$ mean-field effective equations 
\cite{kolomeisky00,salasnich2002,kim03,choi2015} from which 
hydrodynamic equations may be derived as above producing 
(time-dependent) non-linear Schr\"odinger equations
with suitable general non-linear terms $\epsilon(\rho)$, the GP 
equation corresponding to $\epsilon(\rho) \propto \rho$. As discussed 
in \cite{franchini2015} a good agreement is found between the hydrodynamic 
results and the GP equation with a power-law $\epsilon(\rho)$. We observe that 
such mean-field equations may fail in correctly describing 
interference between wavepackets, as shown for the Tonks-Girardeau limit 
$\gamma \to \infty$ in \cite{girardeau2000}: in general one expects anyway 
that the mean-field equations work better at short times with respect 
to long times. For instance, the ultimate fate of a soliton configuration due to classical to quantum crossover has been studied in \cite{pustilnik2015}.

\section{The quench Protocol}
\label{sec:quench}

In this Section we discuss in detail the quench protocol. The starting point 
is to prepare the system with a localized excitation. 
In general, since such state cannot be an eigenstate of a translational invariant system, in time it would diffuse and disperse. However, it is an empirical 
observation that several systems abruptly taken away 
from equilibrium, settle back in configurations displaying 
localized excitations that propagates for long times without 
degrading significantly. 
Such behavior is that of a soliton (or trains of solitons) 
and it is a manifestation 
of the emergent collective hydrodynamics ensuing in the system. 
True solitons are a characteristic feature 
only of integrable differential equations \cite{faddeev}. 
Nonetheless, individual (or well separated) solitonic waves are 
commonly observed in a variety of systems, including of course 
classical systems \cite{newell} 
and ultracold systems \cite{kevrekidis}.

For the quantum many-body Lieb-Liniger model of interacting 
bosons in $1D$ it is not obvious what is the quantum content 
of the microscopical state that results into a soliton. 
It makes sense that it would contain a large number of eigenstates 
such that the coherent dynamics prevents their macroscopic spreading, 
similarly to what happens in quadratic theories for coherent states. 
Indeed, the construction of quantum states with one particle density 
evolving in a solitonic way has not provided conclusive results so far 
\cite{streltsov2011,kaminishi2013,delande2014}, in the sense that it is not 
clear what is the definition of a quantum dark soliton from the Bethe ansatz 
solution, and if the quantum dark soliton may be defined at all (see however 
the very recent discussion in \cite{sato2016}). 

In this work we do not try to solve this problem, but we hope 
that our analysis can contribute 
in this direction. 
Our starting point is the empirical observation that solitonic configurations are commonly excited (and manipulated) in cold atomic Bose 
\cite{kevrekidis,burger1999,khaykovich2002,lamporesi2013} 
and Fermi \cite{yefsah13} systems and that they are best 
understood in terms of an effective semi-classical hydrodynamical 
description of the system. 

Once a solitonic excitation is created, its density profile will remain 
(approximately) constant and will only translate in space at a 
constant velocity. In our quench protocol, at some 
time during the soliton evolution 
we change the parameter $g$ governing the interaction strength 
of the system to $g'$. This interaction quench modifies 
the hydrodynamic equations so that the initial soliton 
cannot evolve unperturbed anymore. We aim at describing 
the dynamics, and in particular the short time dynamics, 
of the soliton after the quench.

\begin{figure}
	\begin{center}
		\includegraphics[width=\columnwidth]{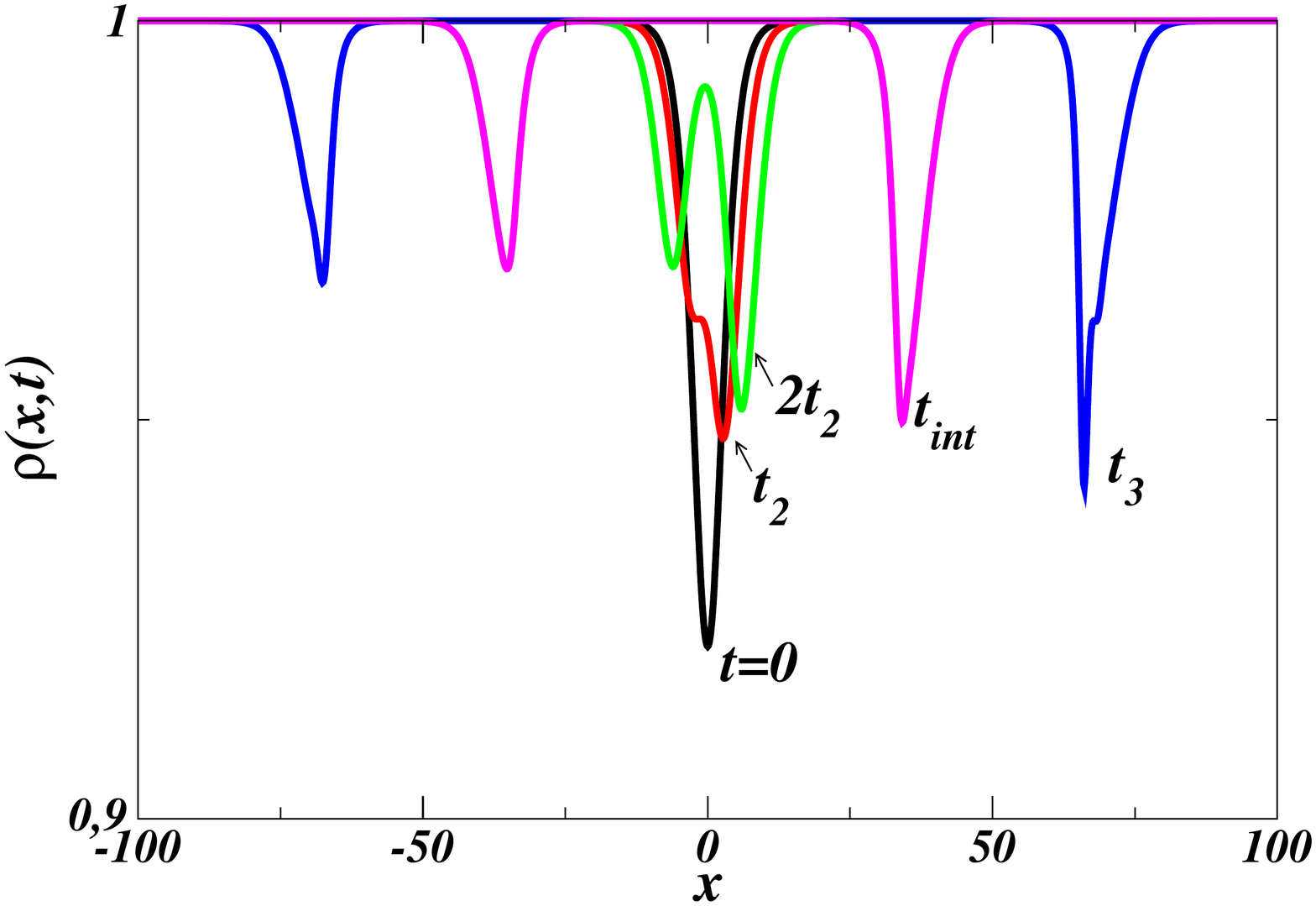}
		\caption{Plot of the density $|\psi(x,t)|^2$ at $5$ different times 
			with $g'/g=20$ and initial velocity $V=0.96c$ 
			(with $g=1$ and $\hbar=m=\rho_0=1$). The interaction is suddenly changed at $t_Q=2 \cdot 10^{-4}$ very close to zero. 
			The black line is the density at $t=0$, and the others are at the times $t=t_2$ (in which 
			a second minimum emerges) - red; $t=2t_2$ - green; $t=t_{int}$ - magenta; $t=t_3$ (in which another minimum 
			is seen in the transmitted packet) - blue. Numerical values are $t_2 \approx 0.708$, $t_{int} \approx 8.219$ and 
			$t_3 \approx 15.729$.}  
		\label{fig:evolution}
	\end{center}
\end{figure}

\begin{figure}
\begin{center}
\includegraphics[width=\columnwidth]{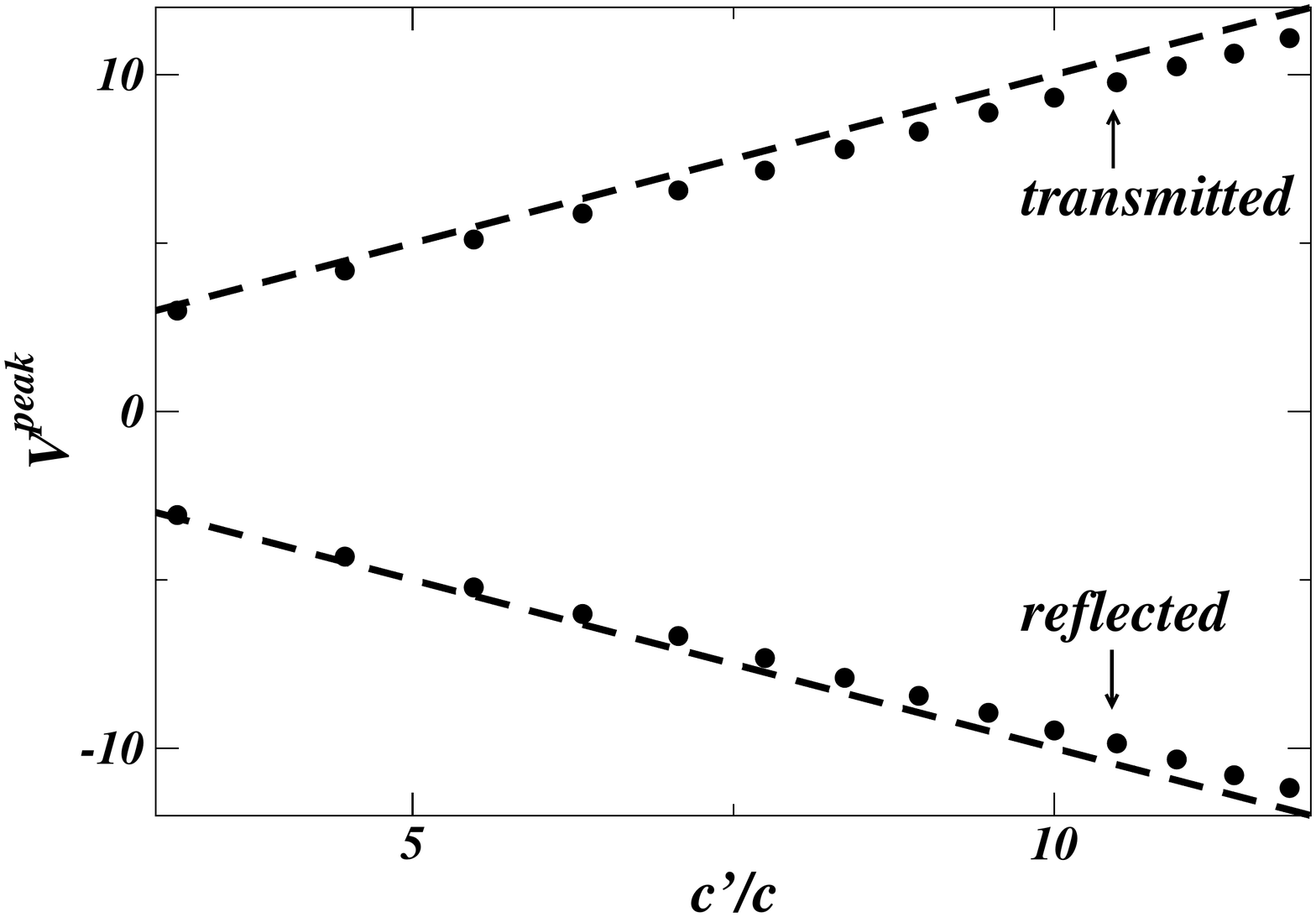}
\caption{Transmitted and reflected peak velocities as a function of $c'/c$ for the same parameters of 
Fig. \ref{fig:evolution}. Dashed lines are the linear predictions $\pm c'/c$ respectively 
for the transmitted and reflected velocities. These velocities and the quantities plotted in figures \ref{BulkVSPeak}, \ref{R_and_T}, \ref{calogeropeakt}, \ref{RT_Cal} are measured at $t_{int}$, intermediate between the time at which the transmitted and reflected profiles are first distinguishable and the time at which the transmitted profile further splits into two.}  
\label{fig:V}
\end{center}
\end{figure}

To qualitatively illustrate what happens after an interaction quench, we take as an example  
the GP equation (\ref{NLS}) and we change the interaction parameter $g$ to a value $g'>g$ 
at the time $t_Q$. The sound velocity changes from $c=\sqrt{g}$ to $c'=\sqrt{g'}$ (in units where 
$\hbar=m=\rho_0=1$ where $\rho_0$ is the density for $x\to\infty$). As initial 
condition we take the gray soliton with velocity $V$ \cite{stringari,novikov1984} reported in Appendix \ref{app:NLSE}
\begin{equation}
\psi(x,t=0)=i\,\frac{V}{c} + \gamma \tanh{\gamma c x},
\end{equation}
where $\gamma=\sqrt{1-V^2/c^2}$. As illustrated in Fig. \ref{fig:evolution} there is a time 
$t_2$ at which the soliton splits in two (more precisely: at which 
a second minimum is seen), and a time $t_3$ in which another splits occurs and one 
of the two packets in turn splits in two as well. We also define an intermediate 
time $t_{int}=(t_2+t_3)/2$. The numerical values of the transmitted and reflected velocities 
are plotted in Fig. \ref{fig:V}: these numerical values are obtained by computing the 
velocities of the (transmitted and reflected) peaks from the peak positions around $t_{int}$, which reduces 
numerical fluctuations 
(as we discuss later and show in Fig. \ref{fig:peak_vel},
we verified that between $t_2$ and $t_3$ the peak velocities are rather stable, 
while close to $t_2$ and $t_3$ numerical fluctuations may be present). 

In Fig. \ref{fig:V} we plot the analytical prediction $\pm c'/c$ obtained from a linear approximation 
(as discussed below), where the $+$ ($-$) is for the transmitted (reflected)
components. The hydrodynamic approach we introduce in the following reproduces 
the leading behavior and it is found to be able to reproduce deviations from it. It is also possible 
to use the hydrodynamic approach to discriminate the bulk and peak velocities: indeed, from the hydrodynamic 
theory one has that for very short times after the interaction (well before $t_2$ where a second minimum 
is observable) the transmitted and reflected packet move as transmitted and reflected solitons. The numerical 
findings we present in the following confirm the validity of the hydrodynamic results up to the time scale $t_{int}$. 

To set up the hydrodynamic approach for the propagation 
of the localized excitations after the interaction quench, as we argued above 
we take the dynamics before the quench to be captured by 
the equations (\ref{cont},\ref{Euler}), where the hydrodynamic parameters 
$\omega(\rho;g)$ and $A(\rho;g)$ depend on a microscopical parameter $g$, setting the strength of the inter-particle interaction. The shape of the initial soliton is set by the coupling $g$ and by its velocity $V$. The approach we develop 
in this section is independent from the microscopic details and it applies 
to general $1D$ systems with solitonic excitations, and it will be compared with numerical results 
in the next sections.

The existence of a soliton is a strong indication 
that it is possible to isolate out of (\ref{cont},\ref{Euler}) 
an integrable core, with the remaining terms being 
negligible for a relatively long time. If the initial 
soliton has velocity $V$ close to the speed of sound $c$ 
(so that its amplitude is also small compared to the background density $\rho_0$), it is known that the hydrodynamics can be reduced to that of the integrable 
KdV. We outline the reduction of equations (\ref{cont},\ref{Euler}) 
to KdV in Appendix \ref{app:KdVreduction}. 
This procedure, developed for cold atoms in \cite{kulkarni2012}, 
is at the heart of our analysis and thus in the following we assume ${V \over c} \simeq 1$. 

The KdV is a chiral equation that reads
\be
  \label{2KdV}
  \dot{u}_\pm \mp \p_x \left[ cu_\pm + {\zeta \over 2} u_\pm^2 - \alpha \p_x^2 u_\pm \right] = 0 \; ,
\ee
where $c \equiv \sqrt{ \rho_0 \omega'_0}$ is the sound velocity ($\omega'_0 \equiv \partial_\rho  \omega|_{\rho_0}$) and the nonlinear and dispersive coefficients are given by
\be
  \zeta \equiv  \frac{c}{\rho_0} + \frac{\p c}{\p \rho_0} \; , \qquad
  \alpha \equiv \frac{A (\rho_0) }{4c} \; . 
  \label{zetaalphadef} 
\ee
Details on the derivation of the KdV equation in 
a generic hydrodynamic system are given in 
Appendix \ref{app:KdVreduction}, while in 
Appendix \ref{app:KdV} we collect some basic results on the KdV. 

The $\pm$ in (\ref{2KdV}) refers to the two chiralities of the waves while $u(x,t)$ is approximately the density fluctuation over the background $u \simeq \rho - \rho_0$.
The KdV reduction neglects the interaction between the left and right moving sectors. Due to locality, such approximation is violated only when the two chiral profiles overlap. However, since they move with relative velocity of approximately $2c$, such effects exists only for short times and hence can be neglected \cite{novikov1984}.

Suddenly during the evolution of the soliton, we change the interparticle coupling to $g'$. Thus, after the quench the hydrodynamic equations will change and the effective dynamics will be given by the KdV (\ref{2KdV}) with modified parameters $\zeta'$, $\alpha'$ and a new speed of sound $c'$. This sudden change in the interaction is seen by the soliton profile as a perturbation to which it reacts by splitting into a transmitted and reflected profile, exactly as it would happen to a linear wave, in the presence of an obstacle, or if the sound velocity is suddenly changed. 

Before the quench, the initial state can be approximated by the KdV soliton (\ref{soldef})
\be
u (x,t) = s (x-Vt) \; .
\ee
We make the ansatz that after the quench 
\be
u (x,t) = u_r (x - V_r t) + u_t (x - V_t t) \; ,
\ee
that is, we assume that the quench acts as an external force which splits the soliton into a transmitted and a reflected profile and that, for short times after the quench, these bumps evolve with a given velocity, without changing their shape significantly. If  $V > 0$, the reflected velocity $V_r < 0 $ and the transmitted one is $V_t  > 0$. 
Imposing continuity of the solution and conservation of momentum at $t=0$ we obtain
\be
u(x,t) = R(V, V_r, V_t) \:  s (x - V_r t) + T(V, V_r, V_t) s (x - V_t t) \; ,
\label{usol}
\ee
with
\be
R (V, V_r, V_t) = {V_t - V \over V_t - V_r}  \; , \qquad {\rm and} \qquad
T (V, V_r, V_t) = {V - V_r \over V_t - V_r} \; .
\label{RTdef}
\ee

Note that, applying the same quench protocol to an initial sound wave moving at the pre-quench speed of sound $c$ would yield transmitted and reflected waves moving at the post-quench sound speed $\pm c'$. The above formulae specialize for this case to
\be
R_{\rm linear} = {c' - c \over 2 \: c'} = {1 \over 2} \left[ 1 - {c \over c'} \right] \; , 
\qquad {\rm and} \qquad
T_{\rm linear} = {c' + c \over 2 \: c'} = {1 \over 2} \left[ 1 + {c \over c'} \right] \; .
\label{RTlinear}
\ee

The derivation of (\ref{usol}, \ref{RTdef}) is completely general, based only on the ansatz that after the quench the soliton splits into two chiral profiles, and does not depend on the dynamics. The laws governing the evolutions are needed to determine the velocities of the two profiles.

We can estimate the profiles velocities by looking at the motion of their center of mass\footnote{We thank the Referee for suggesting this approach.}
\be
   \langle x \rangle_u = {\int x \: u d x \over \int u d x} \; .
\ee	
The velocities are thus
\be
	V_{r,t} = \partial_t \langle x \rangle_{u_{r,t}} = {1 \over \int u_{r,t} \: d x} \:
	\int x \: \dot{u}_{r,t} \: d x \; ,
\ee
where we used the fact that the denominator is the integral of motion $I_0$ (\ref{ndef}) and thus does not evolve with time. We can trade the time derivative for a spacial one through the post-quench equation of motion (\ref{2KdV}):
\be
   V_{r,t} = \pm
   {1 \over \int u_{r,t} \: d x} \: \int x \: 
   \p_x \left[ c' u_{r,t} + {\zeta' \over 2} u_{r,t}^2 - \alpha' \p_x^2 u_{r,t} \right] 
   d x \; ,
\ee
where the $+$ ($-$) sign applies to the left (right) moving profile. We can now integrate by part (remembering that at large distances the profiles decay to zero exponentially) to get
\be
  V_{r,t} = 
  \mp {\int \left[ c' u_{r,t} + {\zeta' \over 2} u_{r,t}^2 \right] d x 
  \over \int u_{r,t} \: d x} 
  = \mp \left[ c' +  {\zeta' \over 2} \: {\int u_{r,t}^2 d x 
  	\over \int u_{r,t} \: d x} \right] \; .
\ee
Note that this expression has a natural interpretation as the ratio between the first two integrals of motion (\ref{ndef}, \ref{I1def}), which can be interpreted as the mass and momentum. Indeed, in (\ref{deltaVest}) we show that this is the case for a soliton:
\be
c- V = {I_1 \over I_0}
= {\zeta \over 2} \: {\int s^2 d x \over \int |s| d x} \ ; .
\label{vest}
\ee
The integral of motions of the two profiles after the quench are determined immediately at their creation and, remembering (\ref{usol}), are simple rescaling of the original ones:
\bea
   V_r & = & - c' + {\zeta' \over 2}  {\int u_r^2  d x \over \int |u_r| d x} 
   = - c' + R \: {\zeta' \over 2}  {\int s^2  d x \over \int |s| d x}   \; , \\
   V_t & = & \: \: c' - {\zeta' \over 2} {\int u_t^2 d x \over \int |u_t| d x}
   = \: \: c' - T \: {\zeta' \over 2} {\int s^2 d x \over \int |s| d x}   \; .
\label{VrtSyst}
\eea
We can now use (\ref{vest}) and (\ref{RTdef}) to get
\bea
  V_r & = & -c' + {\zeta' \over \zeta}  {V_t - V \over V_t - V_r} (c - V) \; , \\
  V_t & = & \: \: c' -{\zeta' \over \zeta}  {V - V_r \over V_t - V_r} (c - V) \; .
\eea
Solving this system we finally get
\bea
V_r & = & - \left[ c - \eta \:  R \: (c - V) \right] {c' \over c} \; ,
\label{dVcmR} \\
V_t & =& \quad \left[  c  - \eta \: T \: (c - V)  \right] {c' \over c} \; , 
\label{dVcmT}
\eea
where we introduced the universal parameter
\be
\eta \equiv {c \over c'} \: {\zeta' \over \zeta} 
= {1 + {\rho_0 \over c'} \: {\partial c' \over \partial \rho_0} \over
	1 + {\rho_0 \over c} \: {\partial c \over \partial \rho_0} } \; ,
\label{etadef}
\ee
and where the $T$ and $R$ are found consistently to be 
\bea
R & =&  {1 \over 2} \left[ 1 - {c \over c'} \: { V \over \eta \:  V  + (1 - \eta) \: c  } \right] \; ,
\label{Rest} \\
T & = & {1 \over 2} \left[ 1 + {c \over c'} \: { V \over \eta \:  V  + (1 - \eta) \: c  } \right] \; .
\label{Test}
\eea
We note that these expressions are completely universal. 
Eqs. (\ref{dVcmR}--\ref{Test}) already appeared in \cite{franchini2015} 
with a different, more qualitative, derivation.

To further highlight the universality of Eqs. (\ref{dVcmR}--\ref{Test}), 
we can make everything dimensionless, by measuring velocities in units of sound velocity. Thus, we introduce the {\it reduced velocities}:
\be
\nu    \equiv {c  -  V \over c} \; , \qquad 
\nu_r \equiv {c' + V_r \over c'} \; , \qquad {\rm and} \qquad
\nu_t \equiv {c'  -  V_t \over c'} \; ,
\label{redv}
\ee
and we write (\ref{dVcmR},\ref{dVcmT}) as
\bea
\nu_r & = & \eta \:  {\nu \over 2} \left[ 1 - {c \over c'} \: 
{1 - \nu \over 1 - \eta \: \nu } \right] \; ,
\label{dVLcm} \\
\nu_t & = & \eta \: {\nu \over 2} \left[ 1 + {c \over c'} \: 
{1 - \nu \over 1 - \eta \: \nu} \right] \; , 
\label{dVRcm}
\eea

Hence, the solution immediately after the quench is
\bea
u (x,t) & = & \: \: \:
 {1 \over 2} \left[ 1 - {c \over c'} {1 - \nu \over 1 - \eta \: \nu} \right] 
s \left[ x + c'(1 - \nu_r) t \right]
\nonumber \\
&&+ {1 \over 2} \left[ 1 + {c \over c'}  {1 - \nu \over 1 - \eta \: \nu} \right]  
s \left[ x - c'(1 - \nu_t) t \right] \; .
\eea
The reflection and transmission coefficients are 
\be
R =  {1 \over 2} \left[ 1 - {c \over c'} {1 - \nu \over 1 - \eta \: \nu} \right] \; , \qquad
T =  {1 \over 2} \left[ 1 + {c \over c'} {1 - \nu \over 1 - \eta \: \nu} \right] \; ,
\label{RTest}
\ee
and the height of each chiral profile is
\be
U^r = R U \; , \qquad U^t = T U \; ,
\label{URTest}
\ee
where $U$ is the height of the pre-quench soliton.

Since the chiral profiles are not solitons of the post quench dynamics, their shape will change during the evolution, because different parts of each profile will move at different speeds. 
For short times after the quench, we can take advantage of the fact that each of the reflected and transmitted profiles are similar to the original soliton, just with a reduced height. Thus, we can use (\ref{Vofx}) to estimate the velocities of the different parts:
\bea
V_r & = &  -c' + {\alpha' \over \alpha} \: c \nu 
-  \left( \zeta' - {\zeta \over R} \: {\alpha' \over \alpha}  \right) u_r(x)  \; , \\
V_t & = & \: \: \: c'   - {\alpha' \over \alpha} \: c \nu 
+  \left( \zeta' - {\zeta \over T} \: {\alpha' \over \alpha}  \right) u_t(x) \; ,
\eea
where we used the fact that the width of both profiles is $W = 2 \sqrt{\alpha \over c \nu}$ and their heights are $U_r = {3 \over \zeta} R c \nu$ and $U_t = {3 \over \zeta} T c \nu$.
As a consistency check, we notice that the velocity at the average height of each profile ($\langle u_{r,t} \rangle = {1 \over 3} U_{r,t}$) coincides with (\ref{dVLcm}, \ref{dVRcm}).
The velocities of the profile peaks are
\bea
V_r^{\rm Peak} & = & -c' + 3{ \zeta' \over \zeta} R c \nu -2 {\alpha' \over \alpha} \: c \nu  \; , \\ 
V_t^{\rm Peak} & = & \: \: \: c ' - 3 {\zeta' \over \zeta} T c \nu  +  2 {\alpha' \over \alpha} \: c \nu \; ,
\eea
which, in terms of the reduced velocities (\ref{redv}) are
\bea
\nu_r^{\rm Peak} & = &
3 \: \nu_r - 2 {\eta \over \beta} \: \nu
= \left[ 3 -  {2 \over \beta_r} \right] \nu_r \; ,
\label{dVLpeak}\\
\nu_t^{\rm Peak} & = & 
3 \: \nu_t  - 2 {\eta \over \beta} \: \nu\ 
= \left[ 3 - {2 \over \beta_t} \right] \nu_t \; ,
\label{dVRpeak} 
\eea
where
\be
\beta \equiv \frac{\Omega_{\zeta'}}{\Omega_{\alpha'}} = {\zeta' \over \alpha'} U W^2 
\simeq \frac{ \zeta' }{\zeta} \: \frac{\alpha}{ \alpha' } \; ,
\label{betadef}
\ee
and
\be
 \beta_r \equiv R \: \beta  \; , \qquad {\rm and} \qquad
 \beta_t \equiv T \: \beta \; .
 \label{betartdef}
\ee
The parameters $\beta_{r,t}$ characterizes whether the dynamics of the chiral profiles is dominated by the non-linear term of the KdV ($\beta_{r,t} \gg 1$), by the dispersive term ($\beta_{r,t} \ll 1$), or is in the solitonic regime of equilibrium between the two ($\beta_{r,t} \simeq 1$).
Eq. (\ref{betadef}) is obtained considering that before the quench the dimensionless ratio $\frac{\Omega_{\zeta}}{\Omega_{\alpha}}$ in (\ref{dimrat}) is close to unity, since we prepared the initial state in a solitonic state, and thus the soliton parameter satisfy
\be
 {\zeta \over \alpha} \simeq {1 \over U W^2} \; .
 \label{prequenchrat}
\ee

Typically, large quenches $g' \gg g$ gives $\beta \gg 1$ 
(see, for instance, (\ref{betaNLSE})): we see that the peaks have reduced velocities three times bigger than the profile center of mass. For smaller quenches, dispersive effects will reduce the peak speeds and we see that for $\beta_{r,t}<1$, $\nu_{r,t}^{\rm peak} < \nu_{r,m}$. Moreover, for $\beta_{r,t} < {2 \over 3}$, the peak starts moving supersonically and thus we expect the profile to become unstable.

Let us now consider the enthalpy $\omega (\rho)$ in (\ref{Euler}) to be a simple monomial, that is
\be 
\omega (\rho) = \phi(g) \rho^{\kappa-1} \; ,
\ee
then 
\be
c^2 = (\kappa-1) \phi(g) \rho_0^{\kappa-1} \; , \qquad
\zeta = \left( {\kappa + 1 \over 2} \right) {c \over \rho_0} \; , \qquad {\rm and} \qquad
\eta = 1 \; .
\ee
Hence
\bea
\nu_r =  {\nu \over 2} \left[ 1 - {c \over c'} \right] 
= {\nu \over 2} \left[ 1 - \sqrt{\phi(g) \over \phi(g')} \right] \; , & \qquad &
R  =  {1 \over 2} \left[ 1 - {c \over c'} \right]
=  {1 \over 2} \left[ 1 - \sqrt{\phi(g) \over \phi(g')} \right] \; ,
\label{nurmon} \\
\nu_t =  {\nu \over 2} \left[ 1 + {c \over c'} \right] 
=  {\nu \over 2} \left[ 1 + \sqrt{\phi(g) \over \phi(g')} \right] \; , & \qquad &
T =  {1 \over 2} \left[ 1 + {c \over c'} \right]
=  {1 \over 2} \left[ 1 + \sqrt{\phi(g) \over \phi(g')} \right] \; .
\label{nutmon}
\eea
Note that the reflection and transmission coefficients are the same of the linear process (\ref{RTlinear}), although the velocities are not.


\subsection{Quench Protocol for the Gross-Pitaevskii Equation}
\label{sec:NLSquench}

To test the predictions based on the KdV universality, we performed some numerical simulations on physically relevant systems. Cold atomic gases with local interaction are commonly described by the GP equation
\be
  i \hbar \partial_t {  \psi} = \left\{ - {\hbar^2 \over 2 m} \partial_{xx} 
  + f(\rho) \right\} \psi \; ,
  \label{GP}
\ee 
where  $\rho (x,t) = \left| \psi (x,t) \right|^2$ and $f(\rho)=g\rho$. To maintain our discussion 
more general we detail the derivation of the hydrodynamic results for a general $f(\rho)$, but we present 
numerical results for the GP equation (\ref{GP}) with $f(\rho)=g\rho$. Numerical results 
for other choices of $f(\rho)$ and in presence of a trapping potential presented in \cite{franchini2015} 
confirms the general validity of the hydrodynamic results for general GP equation with local interactions. 
\begin{figure}
	\begin{center}
		\includegraphics[width=\columnwidth]{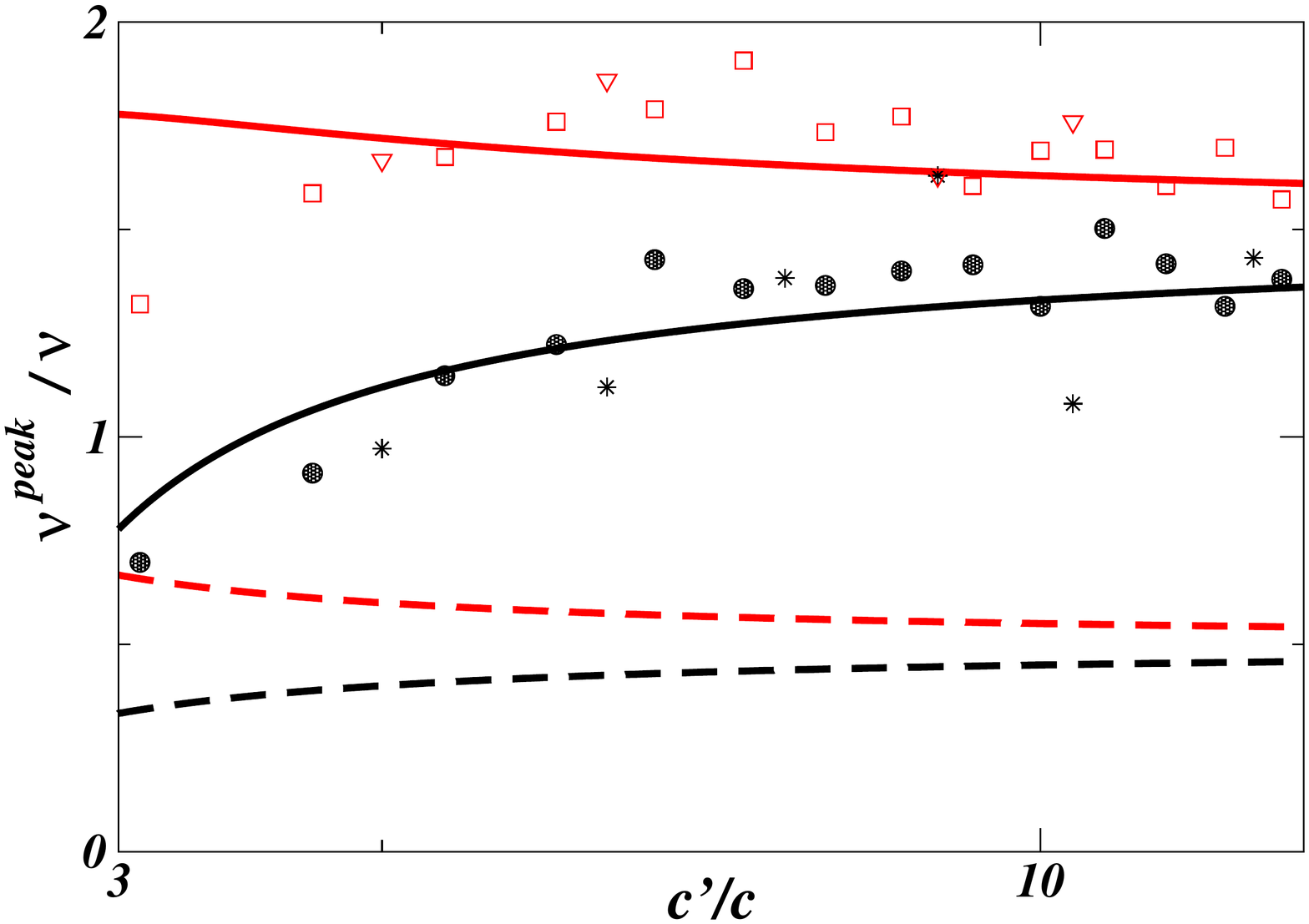}
		\caption{Points are reduced peak velocities in units of $\nu$ {\it vs} $c'/c$ numerically 
			computed from the GP equation: filled black circles ($V=0.96c$) and black stars ($V=0.9c$) 
			are the reflected velocities, empty red squares ($V=0.96c$) and red triangles down ($V=0.9c$) 
			are the transmitted ones. Solid lines are the analytical predictions (\ref{dVR2},\ref{dVL2}) 
			for the reflected (black) and transmitted (red) reduced peak velocities, 
			while the dashed lines are the analytical predictions (\ref{dVL1},\ref{dVR1}) 
			for the reflected (black) and transmitted (red) reduced bulk velocities.}
		\label{BulkVSPeak}
	\end{center}
\end{figure}

In one dimension, the GP equation reduces to (\ref{cont},\ref{Euler}) with the ansatz (\ref{psiansatz}). 
The hydrodynamic functions are
\be 
	\omega(\rho) = \frac{f(\rho)}{m}\; , \quad {\rm and} \quad
	A = \frac{\hbar^2}{2m^2} \:. 
\ee 
	
By substituting the hydrodynamic parameters derived in Appendices \ref{app:NLSE} into \ref{sec:quench} for the GP equation we have
\be
{c' \over c} = {\zeta' \over \zeta} = \sqrt{g' \over g} \; , \quad {\rm and} \quad
\eta = 1 \; .
\ee
Furthermore, remembering the definition (\ref{betadef}) of the parameter $\beta$ we find
\be
  \beta \propto \left( {c^{\prime} \over c} \right)^2 \; .
  \label{betaNLSE}
\ee
Hence, quenching to stronger interactions (higher speed of sound) takes the dynamics to a non-linearity driven regime, as we anticipated. 

Using (\ref{nurmon}, \ref{nutmon}) we get
\bea
\nu_r =  R \: \nu \; , & \qquad &
R = {1 \over 2} \left[ 1 - {c \over c'}\right] \; ,
\label{dVL1} \\
\nu_t =  T \: \nu \; , & \qquad &
T = {1 \over 2} \left[ 1 + {c \over c'}\right] \; .
\label{dVR1} 
\eea

For the peak velocities (\ref{dVLpeak},\ref{dVRpeak}) we have
\bea
\nu_r^{\rm Peak} & = &
3 \: \nu_r - 2 {c^2 \over c^{\prime 2}} \: \nu
= {3 \over 2} \: \nu \left[ 1 - {c \over c'} - {4 \over 3} {c^2 \over c^{\prime 2}} \right] \; ,
\label{dVR2}\\
\nu_t^{\rm Peak} & = & 
3 \: \nu_t  - 2 {c^2 \over c^{\prime 2}} \: \nu\ 
= {3 \over 2} \: \nu \left[ 1 + {c \over c'} - {4 \over 3} {c^2 \over c^{\prime 2}} \right] \; .
\label{dVL2} 
\eea

\begin{figure}
	\begin{center}
		\includegraphics[width=\columnwidth]{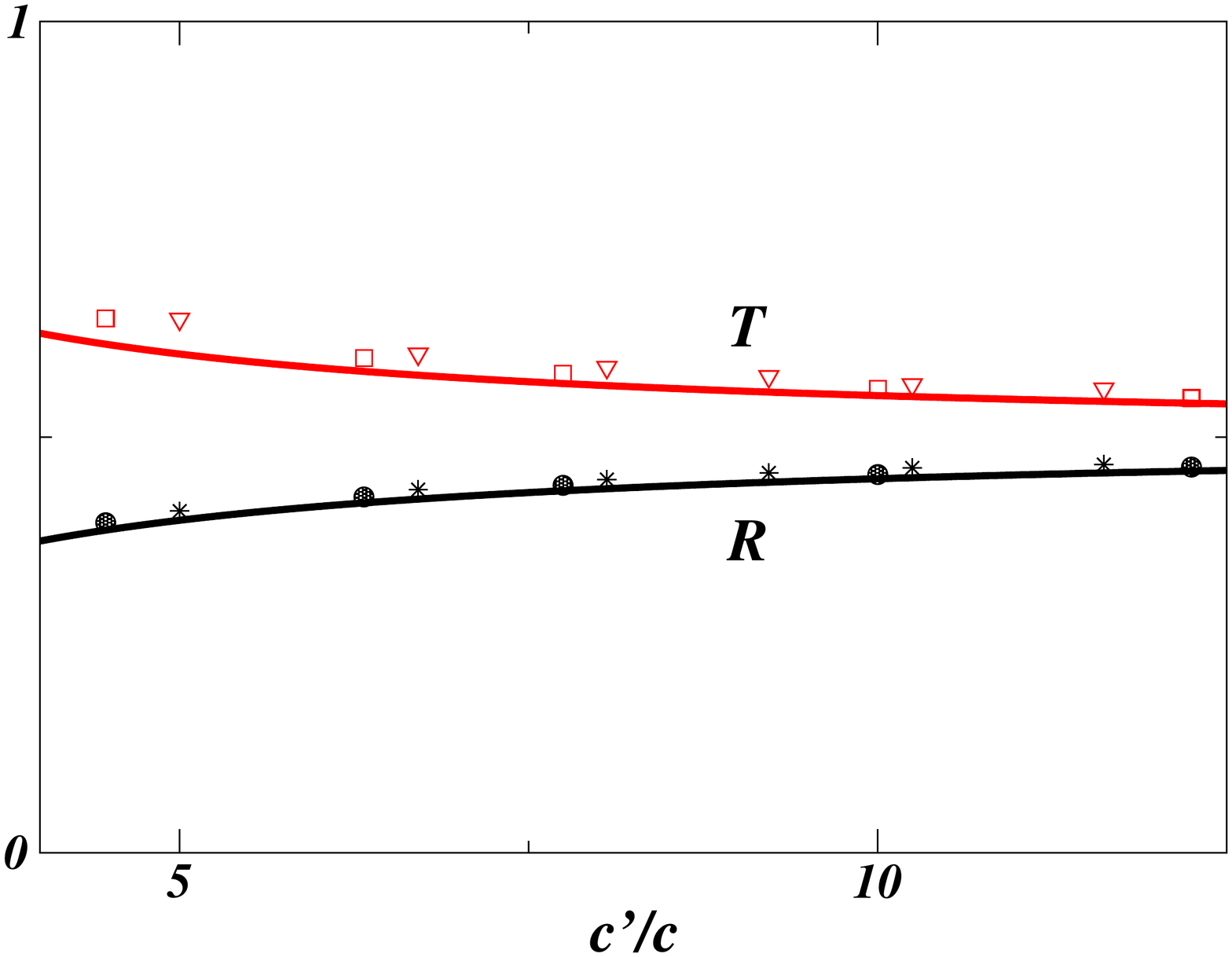}
		\caption{Points are values of $R$ 
			(filled black circles for $V=0.96c$ and black stars for $V=0.9c$) and $T$ 
			(empty red squares for $V=0.96c$ and red triangles down for $V=0.9c$) {\it vs} $c'/c$ numerically 
			computed from the GP equation. Solid lines are the analytical predictions (\ref{dVL1},\ref{dVR1}) 
			for $R$ and $T$.}
		\label{R_and_T}
	\end{center}
\end{figure}

In Fig. \ref{BulkVSPeak} we plot the reduced peak velocities measured during our numerical simulation of the GP equation and compare them to both the prediction for the bulk (\ref{dVL1},\ref{dVR1}) and peak velocities (\ref{dVR2},\ref{dVL2}): it is clearly seen that the numerical results discriminate between the two predictions and 
are rather in agreement with the formulas for the peak velocities. We report in Fig. \ref{BulkVSPeak} numerical 
data for $V=0.96c$ and $V=0.9c$: as expected, numerical data for $V=0.9c$ are more distant 
from the analytical predictions, nevertheless the qualitative properties of analytical 
results (as reduced reflected peak velocities smaller than the transmitted ones and 
both distinct from and larger than bulk velocities) are again visible. 
While in terms of the reduced velocities the behavior of the peak and the bulk is clearly distinguishable, in natural units the two are dominated by the sound velocity drag. This is the reason for which the data plotted in fig. 4 of \cite{franchini2015} is well fitted by the bulk prediction as well.
In Fig. \ref{R_and_T} we plot, 
again for $V=0.96c$ and $V=0.9c$,  
the analytical predictions for $R$ and $T$ from (\ref{dVL1},\ref{dVR1}) and the numerical GP results: 
the agreement is excellent, even better than for the peak velocities. 

\begin{figure}
	\begin{center}
		\includegraphics[width=\columnwidth]{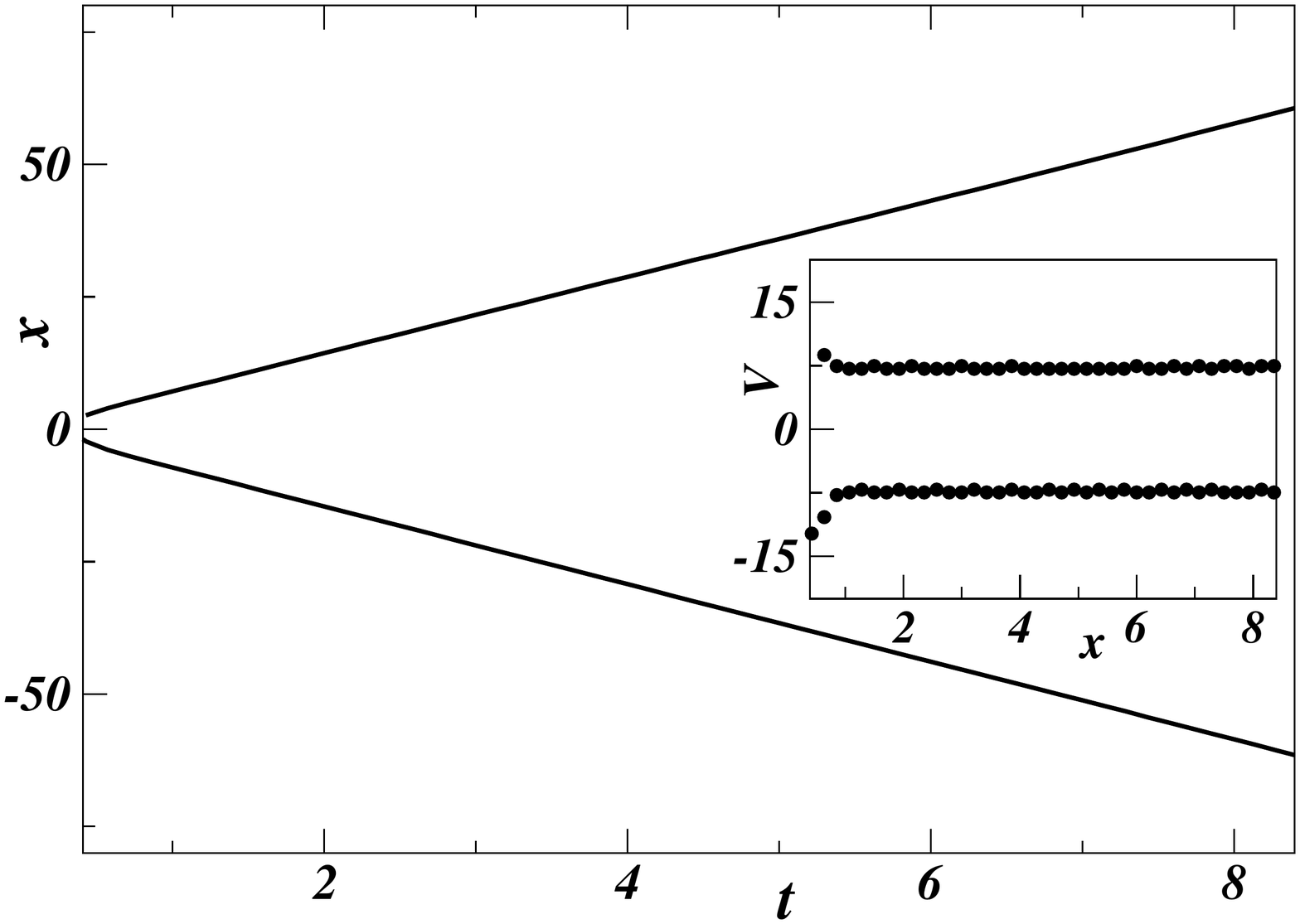}
		\caption{Position of the transmitted (top) and reflected (bottom) peaks 
				for $g'=60$ as a function of time $t$, obtained 
				from the numerical solution of the GP equation, 
				with $t$ between $t_2 \approx 0.407$ 
				and $t_3 \approx 8.400$ ($V/c=0.96$). Inset: numerical estimate 
				of the transmitted (top) and reflected (bottom) peak velocities, 
				as determined by using data spaced in time by $\Delta t=0.21$ and 
				using a spatial grid with $\Delta x=0.07$.}
		\label{fig:peak_vel}
	\end{center}
\end{figure}

The comparison of Figs. \ref{BulkVSPeak} and \ref{R_and_T} shows that the hydrodynamic approach gives good results for both bulk and peak velocities, and that the theory can discriminate between them. 
The data in these figures were taken at time $t_{int}$ defined as halfway between the time $t_2$ were the two profile become distinguishable after the quench and the time $t_3$ where the transmitted profile further splits in two. The dynamics around $t_3$ is clearly beyond our approximation scheme and its treatment requires a different approach, for instance that of \cite{gamayun2014}, which we will not pursue here. Nonetheless, we empirically notice from our numerical experiment that $t_3$ is more than one order of magnitude greater than $t_2$. Moreover, as Fig. \ref{fig:peak_vel} shows, the velocities remain approximately constant in this (large) interval of time.
The possibility of discriminating peak and bulk velocities is important for experiments, where the bulk velocities could be measured with time-of-flight protocols while peak velocities from {\it in situ} imaging of the density. Fig. 
\ref{fig:peak_vel} guarantees that the results of these measurements is not expected to vary significantly between $t_2$ and $t_3$. 
With typical experimental parameters this means that the experiments should 
be order of tens/hundreds on $ms$, which is in turn realistic. 
With the units $\hbar=m=\rho=1$ for $V/c=0.96$ 
we find $t_2 \sim 0.2 - 1$ and 
$t_3 \sim 5 - 20$ for $g' \sim 10 - 100$. If one considers 
a confining frequency $\omega_x \sim 2 \pi \cdot 10 Hz$, 
measuring time in units of $1/\omega_x$ one has that $t_3 \sim 1 - 10$ 
corresponds indeed to to $\sim 10 - 100 ms$. We also observe that a simple 
estimate of $g$ gives with our units 
$g \sim 1$ for a transverse frequency of few $kHz$ and a 
number of particles of order of hundreds (as it is realistic for experiments 
in which one has many $1D$ tubes).

Using (\ref{usol}), we can estimate the time $t_2$ at which it 
is first possible to discern the existence of the two profiles. The comparison 
with numerical results helps to quantitatively 
address the validity of the hydrodynamical approach we are following. 
To do so, we search for the instant at which there is a flex point in (\ref{usol}), that is we look for $(x_2,t_2)$ at which 
the first and second derivative of (\ref{usol}) vanish:
\bea 
  {d u \over d x} & = & {2 \over W} 
  \left[ R \tilde{q} (1-\tilde{q}^2) + T \tilde{p} (1-\tilde{p}^2) \right]= 0 \; 
  \label{1dev}, \\
  {d^2 u \over d x^2} & = & {2 \over W} 
  \left[ R (1-3\tilde{q}^2)(1-\tilde{q}^2) + T (1-3\tilde{p}^2)(1-\tilde{p}^2) \right] 
  = 0
  \label{2dev}
\eea
where we introduced
\bea
  q & \equiv & \tanh \left[ (x-V_r t)/W \right] \; , \\
  p & \equiv & \tanh \left[ (x-V_t t)/W \right] \; ,
\eea
with $V_{r,t}$ given by (\ref{dVL1},\ref{dVR1}) and $W$ by (\ref{NLSEW}).
	
Once we determine the $p$ and $q$ that solves (\ref{1dev},\ref{2dev}), we have 
\be
	t_2 = {W \over V_t - V_r} \left( \arctanh q - \arctanh p \right) \; .
	\label{t2sol1}
\ee

\begin{figure}
	\begin{center}
		\includegraphics[width=\columnwidth]{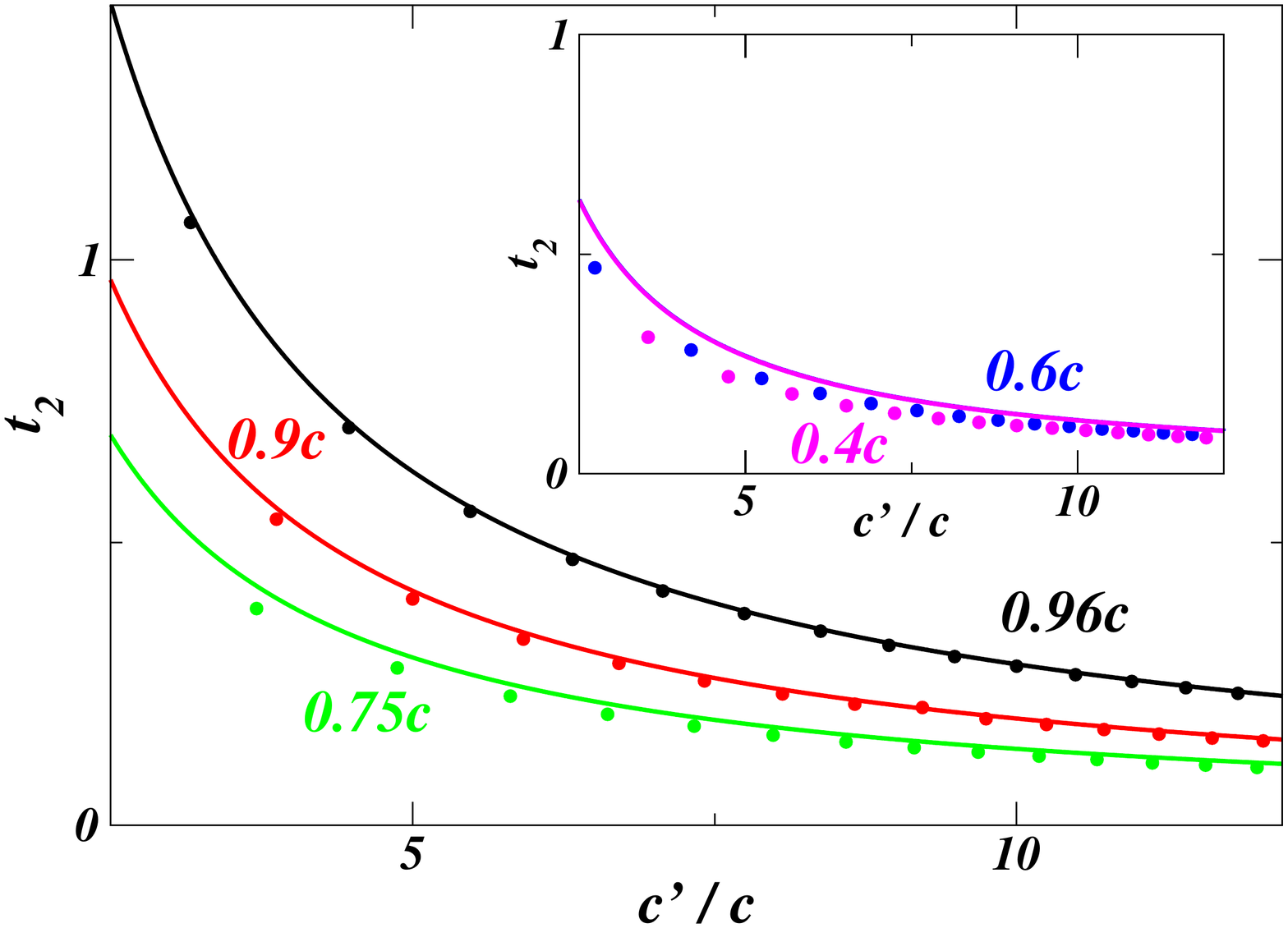}
		\caption{Plot of the time $t_2$ at which the second solitons appears 
				as a function of $c'/c$: Solid lines are the prediction (\ref{t2final}) and 
				dots are obtained from the numerical solution of the GP equation 
				for $V/c=0.96$ (black), $0.9$ (red), $0.75$ (green) from top to the bottom 
				of the figure. Inset: $t_2$ {\it vs} $c'/c$ for $V/c=0.6$ (blue) and 
				$0.4$ (magenta) -- notice that the analytical blue and magenta solid lines 
				are not distinguishable on the scale of the figure.} 
		\label{fig:t2}
	\end{center}
\end{figure}
	
Linear combinations of (\ref{1dev}), (\ref{2dev}) yield the simplified system of equations
\bea
	R q (1-q^2) + T p (1-p^2) & = & 0 \; , 
	\label{syst1} \\
	p (1-3q^2) - q (1-3p^2) & = & 0 \; .
	\label{syst2}
\eea
We can solve (\ref{syst2}), for instance, as a quadratic equation in $q$. One solution is the trivial $q=p$, which corresponds to the equilibrium reached at $x \to \pm \infty$ and is thus not the one we are looking for. The other solution is
\be 
	q = - {1 \over 3 p} \; . 
	\label{qsol}
\ee
Substituting in (\ref{syst1}) we have
\be
  {1 \over 27 p^3} \left[ R - 9 R p^2 +27 T p^4 -27 T p^6 \right] = 0 \; ,
\ee 
which can be solved as a cubic equation in $p^2$. The only real solution is
\bea
	p^2 & = & {1 \over 3} \left\{ 1 + 
	2^{2/3} \left({c' \over c} +1 \right)^{-1/3} \left[ \left({c' \over c} +1 \right) + 
	\sqrt{ {c^{\prime 2} \over c^2} -1} \:  \right]^{-1/3} \right.
	\nonumber \\
	&& \qquad \quad \left. + 2^{1/3} \left({c' \over c} +1 \right)^{-2/3} \left[ \left({c' \over c} +1 \right) + 
	\sqrt{ {c^{\prime 2} \over c^2} -1} \:  \right]^{1/3} \right\} \; ,
	\label{p2sol}
\eea
where we used the definitions of $R,T$ in (\ref{dVL1},\ref{dVR1}) in terms of the quench strength $c' \over c$.

\begin{figure}
	\begin{center}
		\includegraphics[width=\columnwidth]{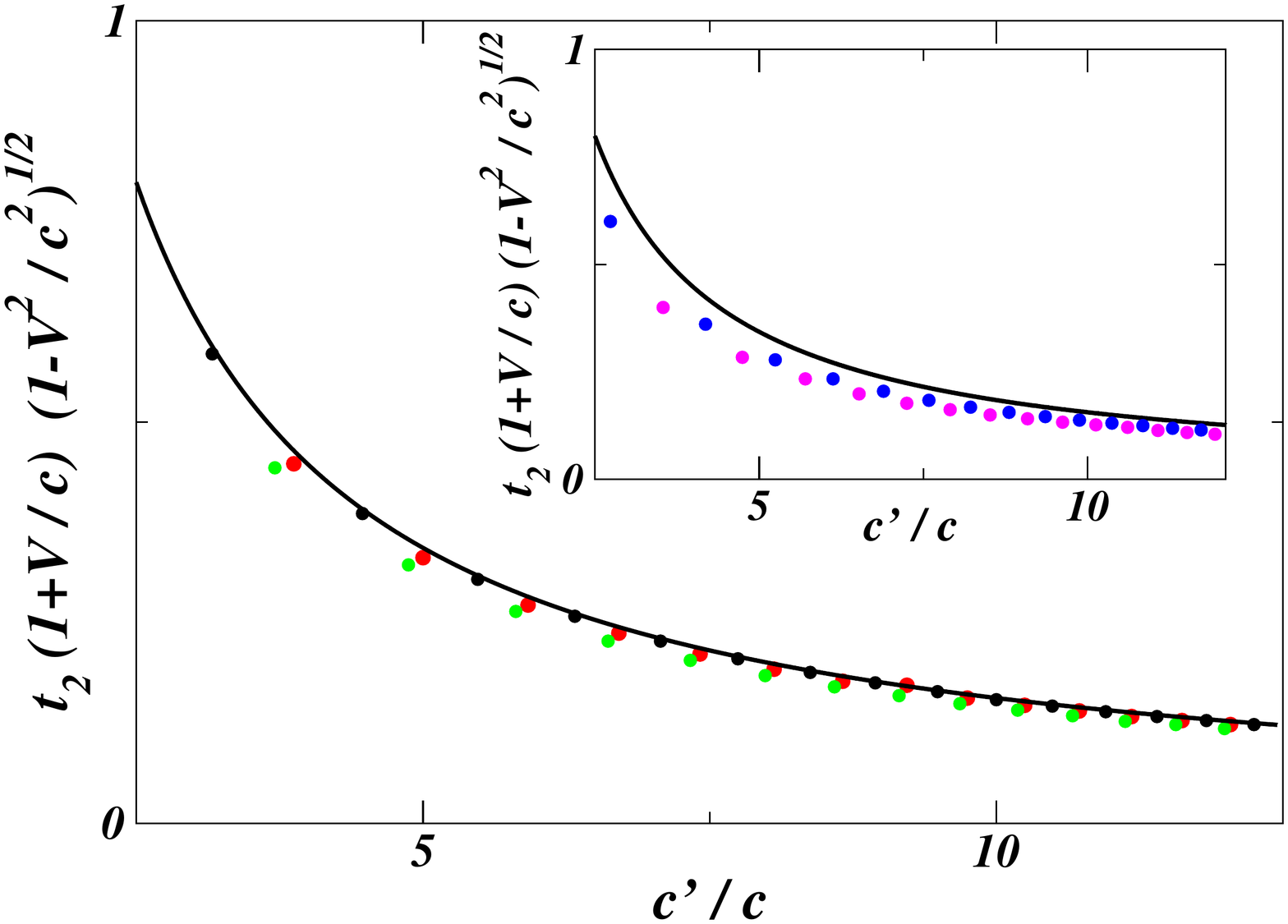}
		\caption{Plot of 
				$t_2$ times $(1+V/c) \cdot \sqrt{1-V^2/c^2}$ as a function 
				of $c'/c$: dots are from the numerical solution of the GP equation 
				for $V/c=0.96$ (black), $0.9$ (red), $0.75$ (green) with $c=1$, while the solid line is the analytical prediction 
				(\ref{t2final_univ}). Inset: same plot 
				for $V/c=0.6$ (blue) and 
				$0.4$ (magenta).} 
		\label{fig:t2_univ}
	\end{center}
\end{figure}

Substituting this solution into (\ref{qsol}) and both of them in (\ref{t2sol1}) and further simplifying the resulting expression we find
\be
	t_2 = {W \over 2( V_t - V_r)}  \ln
	{ \sqrt{ 1 + Q + Q^2} + {\sqrt{3} \over 2} \left( 1 + Q \right) \over
		\sqrt{ 1 + Q + Q^2} - {\sqrt{3} \over 2} \left( 1 + Q \right)  }
	\label{t2final}
\ee
where
\be
   Q \equiv \left( {c' \over c} + \sqrt{ {c^{\prime 2} \over c^2} -1 } \right)^{2/3}\; . 
\ee

We tested this prediction against our numerics, finding an excellent agreement and further supporting the validity of our approximation scheme. The comparison 
is presented in Fig. \ref{fig:t2}. As ths inset of Fig. \ref{fig:t2} shows, 
it is seen that the results are rather good 
also for $V/c$ as low as $0.6$ and $0.4$, where the KdV approximation for the GPE is  not supposed to be reliable. 

\begin{figure}
	\begin{center}
		\includegraphics[width=\columnwidth]{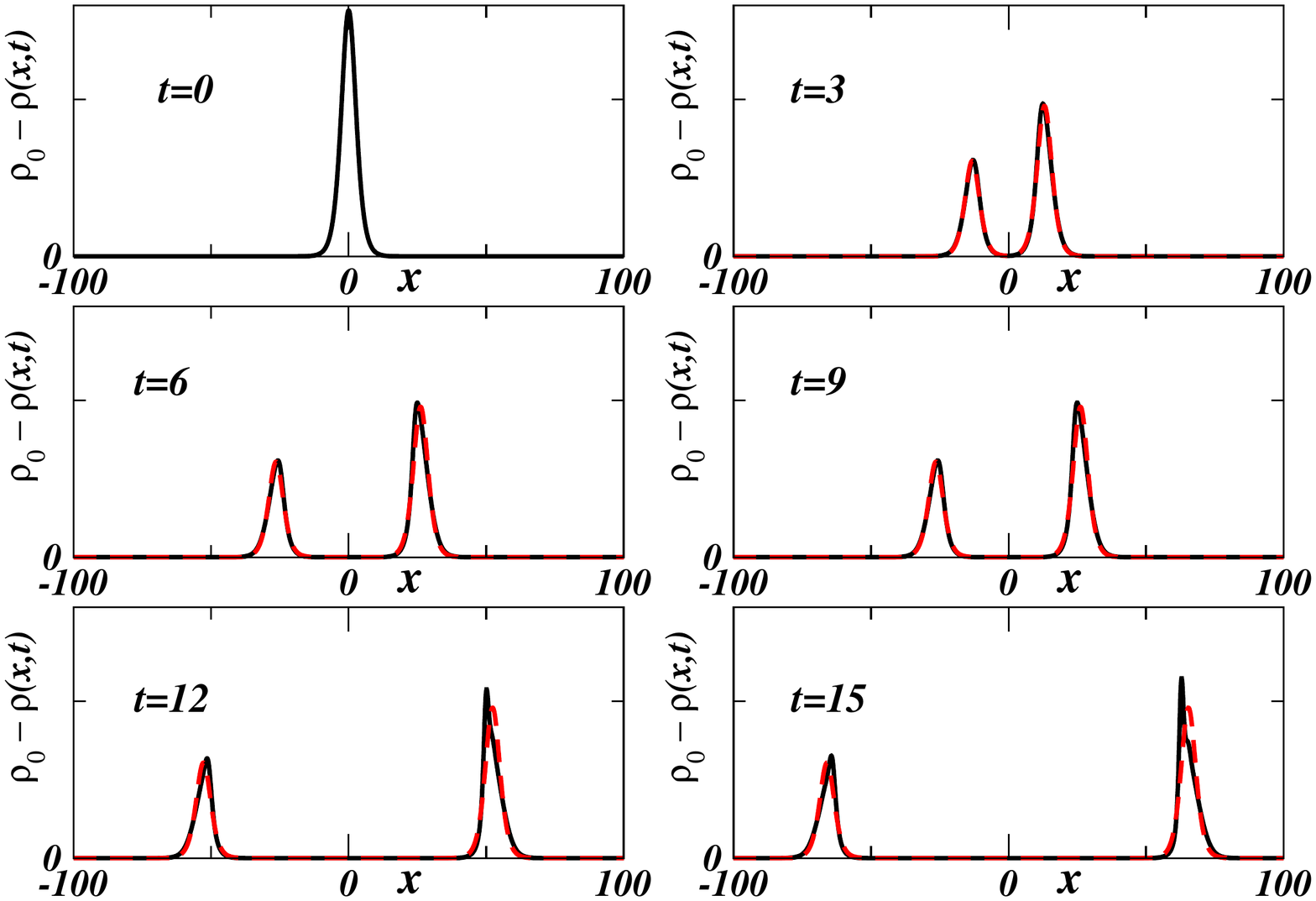}
		\caption{Plot of the density $\rho_0-\rho(x,t)$ of the moving 
				bump for the different times $t=0,3,6,9,12,15$ for $g'=20$ and $V/c=0.96$ 
				(here $t_Q=0$). 
				The solid black lines are numerical results from the GP equation, while 
				the red dashed ones (of course not plotted at the initial time) 
				are from the analytical prediction (\ref{usol}). For the considered parameters 
				it is $t_2\approx 0.728$ and $t_3 \approx 15.749$.} 
		\label{fig:comparison}
	\end{center}
\end{figure}

To quantitatively assess deviations from the prediction (\ref{t2final}) 
we observe that the quantity $t_2$ times $(1+V/c) \cdot \sqrt{1-V^2/c^2}$ 
according Eq. (\ref{t2final}) itself is 
independent from the initial velocity:
\be
	t_2 \left( 1+\frac{V}{c}\right) \sqrt{c^2-V^2} = \frac{1}{2c'} \ln
	{ \sqrt{ 1 + Q + Q^2} + {\sqrt{3} \over 2} \left( 1 + Q \right) \over
		\sqrt{ 1 + Q + Q^2} - {\sqrt{3} \over 2} \left( 1 + Q \right) }.
	\label{t2final_univ}
\ee
Numerical results for $V/c=0.96,0.9,0.75,0.6,0.4$ are reported in 
Fig. \ref{fig:t2_univ}: as expected, one sees that, while for shallow initial solitons the agreement is excellent, by decreasing the initial velocity the agreement becomes less satisfactory, although still acceptable.

To further show how well and for how long the approximations we used remain valid, we plot in Fig. \ref{fig:comparison} the density $\rho_0-\rho(x,t)$ for different times between $t_2$ and $t_3$ from the numerical solution of the GP equation and from the KdV hydrodynamic approach. It is seen that the analytical prediction is rather good almost all the way to $t_3$, even though approaching $t_3$ it cannot reproduce the deformation needed to expel the third soliton.

\subsection{Quench protocol for the Calogero Model}
\label{sec:Calogeroquench}

In the previous section we showed that numerical simulations strongly support our analytical analysis. In this Section we perform an additional check, by employing the Calogero model. Details about the Calogero model, its hydrodynamic 
description and its extension in the presence of an harmonic confinement are 
in Appendices \ref{app:Calogero} and \ref{app:HarmCal} 

The study of the interaction quench in the Calogero model 
allows for to address several different points. 
First, we show that the universality of the quench protocol is robust, 
in that it applies also to long range interaction such as that of Calogero. 
Additionally, we analyze the effect of an external potential and thus of a non-constant background density. Finally, instead of simulating directly the hydrodynamic of the model, we perform a (classical) Newtonian evolution for a system composed of a large number of particles and extract the emerging collective behavior. We will see that the initial soliton will split into a reflected and transmitted density profile, showing an emergent wave behavior out of the individual particle dynamics.

The Calogero model in an external harmonic potential is defined by the Hamiltonian \cite{Calogero-1969,Sutherland-1971}
\be
H = {1 \over 2 m} \sum_{j=1}^N \left( p_j^2 + \omega^2 x_j^2 \right)
+ {\hbar^2 \over 2 m}  \sum_{j \ne k} {\lambda^2  \over (x_j - x_k)^2} \; ,
\label{harmCal}
\ee
with a dimensionless coupling constant $\lambda$. We can perform the numerical evolution in the classical limit [in the quantum Calogero models, the coupling undergoes the quantum shift $\lambda^2 \to \lambda (\lambda -1)$], which is integrable even in the presence of an external parabolic potential. 

This model has a long-range (power-law) interaction and thus it lies in a different universality class, compared to local models. 
At the same time, as was recognized in Ref. \cite{kulNPB},  the potential $1/r^2$
is also relatively short ranged in one dimension (despite being power law). Thus, although such interaction has not yet been realized in cold atomic gases, this model provides a good platform to study systems beyond contact interaction, with the additional benefit of being exactly solvable, even with an external confinement.

The above model (Eq. \ref{harmCal}) has been shown to admit soliton solutions in \cite{abanov2011}. These are very special set of positions and momenta for each particle that collectively evolve as a robust bump on top of a curved density background (see Appendix \ref{app:HarmCal}). 

Due to the power-law interaction, the hydrodynamic description of (\ref{harmCal}) is not the KdV, but a different 
integrable equation, in the family of the Benjamin-Ono (BO) {equation}
{\cite{abanov2009}}.
It differs from the KdV by its dispersive term and by the fact that its 
solitons have longer (power-law) tails. Another difference is that this 
system supports supersonic bright solitons, instead of the subsonic dark 
ones of local models \cite{abanov2011}. We collect in appendix \ref{app:Calogero} and \ref{app:HarmCal} some useful information and new results on the model.

We can calculate the bulk velocities of the reflected and transmitted profiles by operating as before, through the velocity of the center of mass of each. Solving the system of equations (\ref{VrtSyst}) for this case yields the same results as for the KdV,
with $\eta =1$.
To take into account the supersonic nature of the Calogero excitations we define the reduced velocities as
\be
V = c (1 + \nu) \; , \qquad V_r = - c' (1 + \nu_r) \; , \qquad V_t = c' (1 + \nu_t) \; ,
\ee
so that $\nu$ and $\nu_{r,t}$ are all positive. In terms of these, we have
\be
\nu_r = R \nu \; , \qquad \nu_t = T \nu \; ,
\ee
and
\be
\label{req}
R = {1 \over 2} \left[ 1 - {c \over c'}  \right] \; , \qquad
T = {1 \over 2} \left[ 1 + {c \over c'}  \right] \; ,
\ee
which coincides with (\ref{dVL1},\ref{dVR1}).

The solitons in Calogero are Lorentzian (\ref{calsol}) and we notice that their height and width do not depend on the coupling $\lambda$. Hence, if $\beta$ was of the order of unity before the quench, after the quench $\beta_{r,t}$ will be less than unity, proportional to the coefficients $R,T$, due to the height reduction.
Hence, for the Calogero model, after then quench we will be inevitably in the dispersive regime.
We can calculate the peak velocities using (\ref{BOVpeak}), with $a_r = R \rho_0 \tau$, $a_t = T \rho_0 \tau$:
\bea
\nu_r^{\rm Peak} & = & \left[ \zeta' R  - 3 \pi \alpha'_{BO} \right] {\rho_0 \over \tau c'} 
= \left[ 4  - {3 \over R} \right] \nu_r = - \left[ 1 + 2 {c \over c'} \right] \nu \; , \label{nu_c_1}\\
\nu_t^{\rm Peak} & = & \left[ \zeta' T  - 3 \pi \alpha'_{BO} \right] {\rho_0 \over \tau c'} 
= \left[ 4 - {3 \over T} \right] \nu_t = - \left[ 1- 2 {c \over c'}  \right] \nu \; \label{nu_c_2}.
\eea
We see that the reflected peak always move slower than the speed of sound ($\nu_r <0$ in our notations), while the transmitted one remains supersonic only for $c' < 2 c$. 
This results is seemingly counterintuitive, since the profile densities lie above the background that their bulk velocities are supersonics. While, in a non-linearity dominated regime, the profile peaks would move faster than the bulk, we see that effect of the dispersive regime effective after the quench in the Calogero model makes the peaks move even slower that speed of sound.
Thus, while the bulk behavior of the Calogero system after the quench follows the same universality of local models, the peak velocities belong to a different universality.

In our numerical simulation, we used the results of \cite{abanov2011}, where it was shown that a system on $N$ Calogero particle is dual to a system of $M$ interacting complex parameter, where $M$ counts the number of solitons in the system. Thus, to a system of $N$ particle lying on the real axis and interacting through (\ref{harmCal}) we add a dual particle $z(t)$ (see Appendix E) which draws an ellipse on the complex plane. The interaction of the dual variable with the real particles induces a soliton in the latter and the soliton peak follows the projection of the $z$ particle on the real axis. As explained in \cite{abanov2011}, a very useful property of this system is that, given an initial condition for the particle position and momenta, the configuration at any given time $t$ can be found exactly by diagonalizing a certain matrix system (by exploiting the Lax pair formulation of the model). 
In this way, we have been able to set the initial conditions of a soliton, to let it evolve for some time and to follow its evolution after the interaction quench exactly. 

\begin{figure}
	\begin{center}
		\includegraphics[scale=0.4]{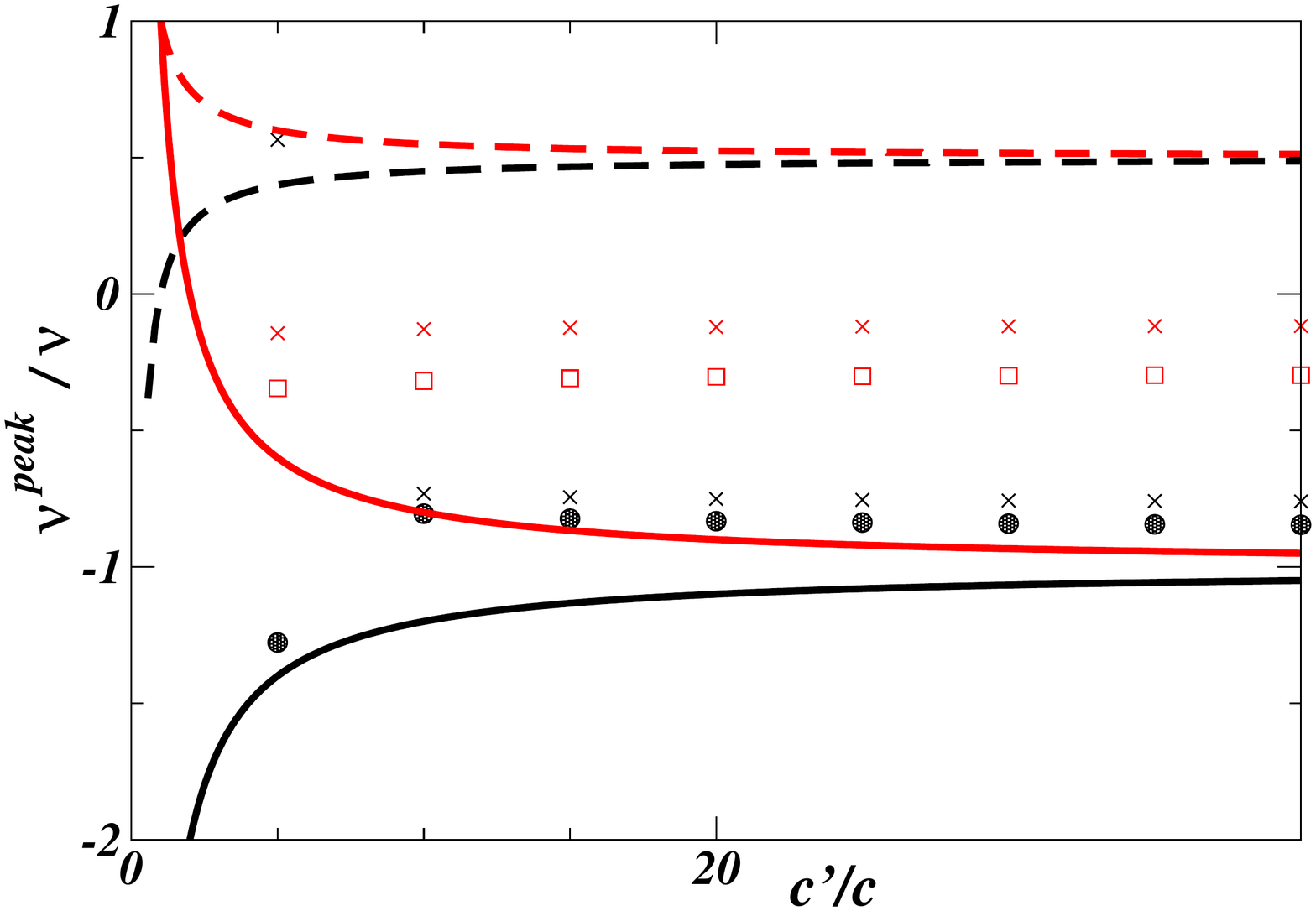}
		\caption{Plot of the reduced peak velocity in units of $\nu$ as obtained from the numerics: 
			filled black circles (black stars) are the values 
			of reflected peak velocities 
			for $\nu=0.04$ ($\nu=0.07$), while empty red squared (red stars) are the corresponding values 
			for transmitted peak velocity for $\nu=0.04$ ($\nu=0.07$). Solid lines corresponds to the analytical 
			predictions (\ref{nu_c_1}) and (\ref{nu_c_2}), respectively for the reflected (black) and transmitted 
			(red) peak velocities. Dashed lines are the predictions for the bulk reflected (black) and transmitted 
			(red) bulk velocities according (\ref{req}).}
		\label{calogeropeakt}
	\end{center}
\end{figure}

In \cite{rajabpour2014} the ground state of the quantum Calogero model in a harmonic potential was studied after a quench and it was observed that the one-particle density starts oscillating and breathing. Such behavior is natural, since after the quench the equilibrium particle distance increases and the trapping cannot compensate for this repulsion. To neutralize this effect, it is thus important to quench at the same time both the interaction and the external potential, so that the background density stays constant. We found that for the model (\ref{harmCal}) the trapping has to be increased by the same amount of the interaction:
\be
   \omega' = {\lambda' \over \lambda} \omega \; .
\ee
Once we are able to stabilize the background in this way, we perform the quench experiment and measure the characteristics of the reflected and transmitted profiles.

To produce the initial soliton configuration, we specify a given complex number $z$ and the initial value problem for the position and momentum of each particle in the system  ($x_j(0)$, $p_j(0)$) is given by the following equations which can be solved numerically:
\bea
	\omega x_j
	&=& \lambda \sum_{k=1 (k\neq j)}^{N} \frac{1}{x_j-x_k}
	-\frac{\lambda}{2} \left( \frac{1}{x_j-z} + \frac{1}{x_j-\bar{z}}\right),
	\la{M1xj} \\
	p_j &=&  i\frac{\lambda}{2} \left( \frac{1}{x_j-z} - \frac{1}{x_j-\bar{z}}\right).
	\la{M1pj}
\eea
Having determined the initial values of $x_j (0)$ and $p_j (0)$, the time dynamics of the system of Calogero particles after the quench can be computed by exploiting the Lax pair formalism, similarly to what was done in \cite{abanov2011} to study the dynamics of a hCM soliton.
One introduces the following $N\times N$ matrices:
\bea
	X_{ij} &=& \delta_{ij}x_i,
	\la{X-matrix} \\
	L^{\pm} &=& L \pm i\omega' X, \;\;\mbox{where } L_{ij} = p_i\delta_{ij} +(1-\delta_{ij})\frac{i \lambda'}{x_i - x_j},
	\la{L-matrix} \\
	M_{ij} &=& \lambda' \left[\delta_{ij}\sum_{l=1 (l\neq i)}^{N}\frac{1}{(x_{i}-x_{l})^{2}}
	-(1-\delta_{ij})\frac{1}{(x_{i}-x_{j})^{2}}\right],
	\la{M-matrix}
\eea
which depend on time through $x_{j}(t)$ and $p_{j}(t)$. It is straightforward to show that the equations of motion, 
\bea
	\dot{x}_{j} &=& p_{j},
	\la{csmeq1} \\
	\dot{p}_{j} &=& -\omega^{\prime 2}x-\lambda^{2} \frac{\partial}{\partial x_{j}}
	\sum_{k=1\, (k\neq j)}^{N}\frac{1}{(x_{j}-x_{k})^{2}}
	\la{csmeq2}
\eea
are equivalent to the following matrix equations
\bea
	\dot{X} +i[M,X] &=& L,
	\la{Xdot-equation} \\
	\dot{L}+i[M,L] &=& -\omega^{\prime 2}X
	\la{Ldot-equation}
\eea
or equivalently
\be
	\dot{L}^{\pm} = -i\left[M,L^{\pm}\right] \pm i\omega' L^{\pm}
	\la{LM-equation}
\ee
written in terms of $L$ and $M$ matrices usually referred to as a Lax pair.
	
One can then write the solution of the hCM as an eigenvalue problem for a matrix which can be explicitly constructed from the initial positions and velocities of the Calogero particles. Namely, the particle trajectories are given by eigenvalues of the following matrix \cite{pere}
\be
	Q(t) = X(0)\cos(\omega' t) + {1 \over \omega'} \: L(0)\sin(\omega' t) \; ,
	\la{Q-explicit}
\ee
where the matrices $X(0)$ and $L(0)$ are constructed using the initial conditions $x_{j}(0)$, $p_{j}(0)$ from (\ref{M1xj}, \ref{M1pj}) inserted in the definitions (\ref{X-matrix},\ref{L-matrix}). It is worth pointing out that the above technique is non-iterative in time and hence there is no numerical error accumulation. 

\begin{figure}
\begin{center}
\includegraphics[scale=0.4]{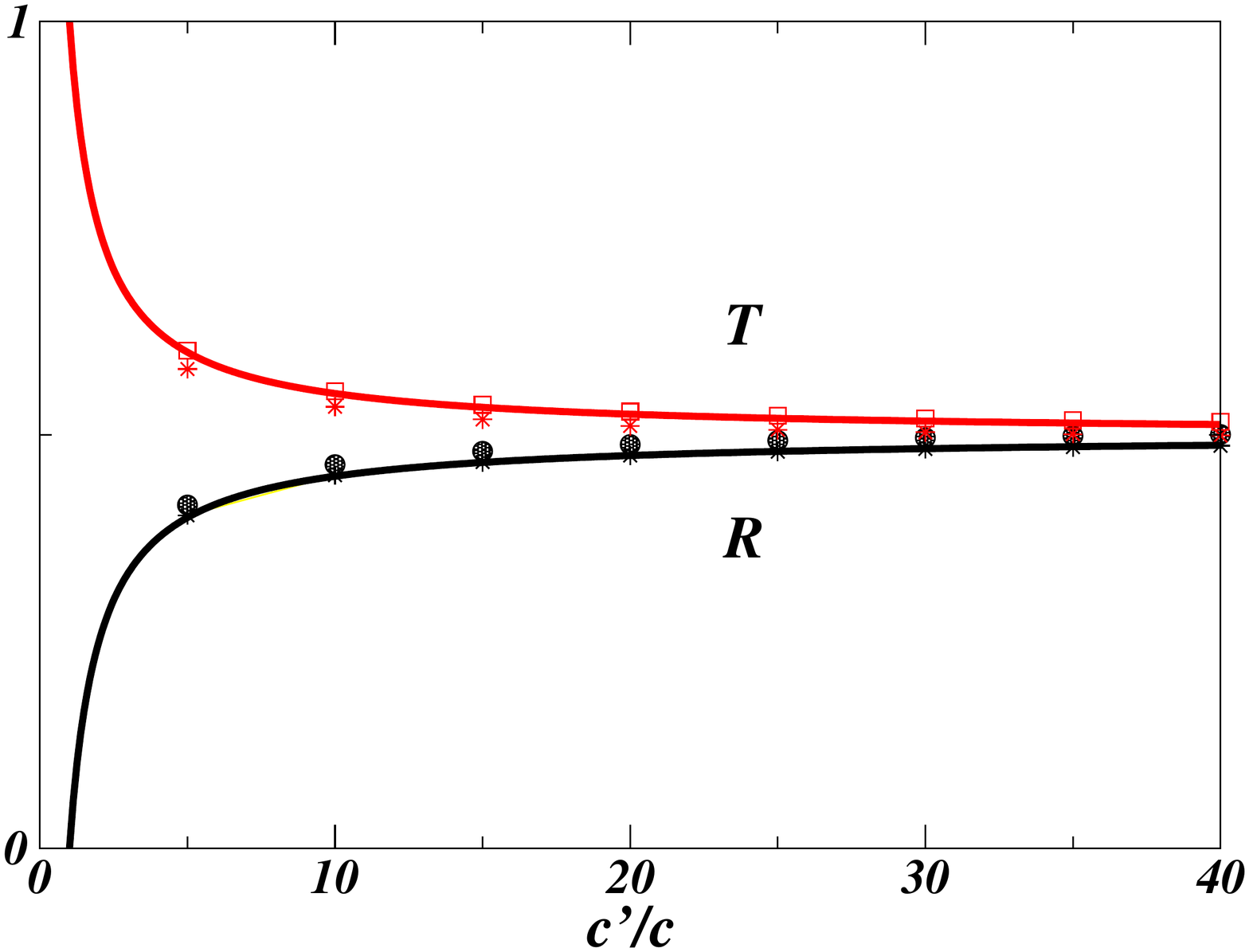}
\caption{Plot of $R$ (bottom black) and $T$ (top red) 
for the quenched Calogero model showing the height of the reflected and 
transmitted peaks as seen from numerics and their comparison with analytical calculations 
(Eq. \ref{req}). Points are numerical results: filled black circles (black stars) are the values 
of $R$ for $\nu=0.04$ ($\nu=0.07$), while empty red squared (red stars) are the corresponding values 
for $T$ again for $\nu=0.04$ ($\nu=0.07$).}  
\label{RT_Cal}
\end{center}
\end{figure}

In our numerical investigations, we used two different initial soliton velocities: one moving $4\%$ faster than the speed of sound and one $7\%$.
In Fig. \ref{calogeropeakt} we present results for the reduced (reflected and transmitted) 
peak velocities for both cases. While the numerical results for the reflected peak velocities are clearly closer to the analytical peak estimates than to the bulk ones, the agreement is less satisfactory for the transmitted peak velocities (the $4\%$ data are closer to the peak velocities, while the $7\%$ data are almost 
in the middle between the peak and the bulk reduced velocities). Nevertheless, these data are clear evidence of the subsonic dynamics predicted by the analytics above, while evidently the peculiarities of the Calogero model renders the quantitative comparison more troublesome. 
In Fig. \ref{RT_Cal}, we show the comparison between analytics (Eq. \ref{req}) and numerics for the reflected and transmitted heights ($R,T$) for the cases of $\nu=4\%$ and $7\%$. A remarkable agreement between these analytical predictions and the numerical calculations performed on the Calogero model is evident.

\section{Discussion \& conclusions}
\label{sec:conclusions}
We have discussed a novel quench protocol, in which a moving, localized excitation reacts to a global interaction quench. We focused on a special class of excitations, namely, soliton solutions, which are stable and experimentally achievable in cold atomic systems as well as in other relevant
interacting low-dimensional systems. We provided a general hydrodynamic framework describing the collective behavior of such systems. Using this hydrodynamic description, we showed that the dynamics immediately after the quench is universal, in that it does not depend on the details of the
microscopic interaction, but only on macroscopic quantities, such a the speed of sound before and after the quench.

The quench protocol can be seen as a initial value problem for the non-linear PDE of hydrodynamic type. This has been the approach of \cite{gamayun2014,gamayun2015,caudrelier2015} and is also common in mathematics. It is thus known that a generic initial condition will
eventually break into several components, each moving with different velocities (see Fig. \ref{fig:evolution}). Under certain conditions, some of these components will be stable, while other (most of them) will disperse. In \cite{gamayun2014} it was noted that, for the integrable non-linear Schr\"odinger, a quench that brings the new speed of sound to
be an integer multiple of the original one will only generate a train of solitons (approximately half of them moving in the same direction as the pre-quench soliton and the other half moving in the opposite direction) and no dispersive sound waves. 

In our approach we focused on short times after the quench and, making no assumptions on the integrability of the models, but using the structure of the initial condition provided by the quench protocol, we predict that the initial excitation will immediately break into two counter-propagating
packets. While they will eventually break further, for a certain time the two chiral packets will retain the shape of the original excitation, but with amplitudes reduced by a reflection and transmission coefficient. The universal form of these coefficients are given in (\ref{Rest},\ref{Test}), while
the bulk velocities of the two profiles appear in (\ref{dVLcm},\ref{dVRcm}). It is also possible to express the latter in dimensionless quantities as in (\ref{dVLcm},\ref{dVRcm}). These velocities are measurable in a time of flight experiments, but in our numerical simulation we employ a direct measurement scheme. Thus, it is simpler to measure the velocities of the peak of the profiles, which differ from the bulk (that is, center of mass) velocities because of the internal redistribution of energies of the packets. The universal expressions for the peak velocities are given in
(\ref{dVLpeak},\ref{dVRpeak}). In Figs. \ref{BulkVSPeak} and \ref{R_and_T} we plot the comparison between the analytical expressions and the numerical results obtained through 1D Gross-Pitaevskii equation, which show a good agreement for the $R$ and $T$ coefficients and an acceptable agreement for the peak velocities. As reported in \cite{franchini2015}, when expressed in natural units, the agreement between the analytical expressions and the numerical data for the velocities is excellent, although it is hard to discriminate between peak and bulk predictions.
We also addressed the experimental feasibility of the proposed protocol: in (\ref{t2final}) we analytically estimated the time at which the two-profiles dynamics becomes discernible and show in figures \ref{fig:t2}, \ref{fig:t2_univ} its agreement with the numerical data. We also argued that measurements can be performed reliably for around $10 - 100 ms$ before intermediate-time effects should be taken into account. Fig. \ref{fig:peak_vel} shows that in this ``short time'' dynamics window the profile velocities stay reasonably constant and Fig. \ref{fig:comparison} that the analytical profiles are well fitted bu the numerical ones.

We also considered the Calogero model. Although we are not aware of cold atomic gases that realize the inverse square interaction of this integrable system, this model allows us to establish several relevant points.
First of all, we see that, despite the non-local nature of this interaction, the short times dynamics remain similar to the local case and we found that the $R$ and $T$ coefficients, as well as the bulk velocities of the chiral profiles, are given by the same universal expression as before (see also \cite{franchini2015}). The peak
velocities, instead, follow a different universality (which is dominated by dispersive effects, instead of the non-linear at play for local interactions). While the solitons and the chiral profiles move supersonically in the Calogero case, the peak velocities are predicted to be subsonic. The agreement
between the analytical expectations and the numerical simulation are less satisfactory in this case, compared to the local interactions, but still support the qualitative behavior derived analytically and the result 
for the subsonic peak velocities.
In our numerical simulation of the Calogero dynamics we employ a classical, Newtonian evolution of the particles and extracted their collective behavior, whose agreement with the hydrodynamic prediction is a further, somewhat independent, proof of the solidity of our approach.

Finally, the Calogero model remains integrable even in the case of an external harmonic potential. Thus, by including the effect of the trap, we established that by simultaneously quenching the interaction between the particles and the external field, it is possible to keep the background density fixed
and isolate the quench dynamics of the soliton. We believe that this observation is applicable to the experimental realization of the protocol and give us 
insights on how to deal with the external trapping.
 
We collected in the Appendices some old and new results concerning the dynamics of solitons in non-linear and cold atom systems. We explained how the dynamics of shallow profiles can always be captured by the KdV equation and in particular how does this work for the GP equation 
(and its generalizations). We also considered the solitons for system of Calogero type and discussed their behavior in the presence of an external trap. In this case, the soliton changes its shape during the evolution, because the background density is also changing. We derived the expression for the
soliton velocity on such background and showed that it decreases moving toward the edges of the systems as the soliton width is also simultaneously decreasing. Such behavior contradicts the common wisdom valid in traditional, translational invariant case of constant background (namely, the
notion that thinner solitons move faster).
 
The realizations of low dimensional bosonic gases along with cutting edge technologies available for their manipulation (e.g., the generation of solitons, the induction of interaction quenches, and the observation of the 
subsequent time dynamics) stimulates the development of theoretical 
tools to interpret these data.
We believe that our work paves the path for new experiments. 
While we assume a cold atomic gas realization as the natural setting 
to implement our quench protocol, we should stress 
that it may be realizable in non-linear optical experiments as well, 
where the non-linear Schr\"odinger equation and non-linear PDE are ubiquitous.

Future directions of investigations include studying the consequences of having multiple local excitations (corresponding to multi-soliton profiles) and their subsequent behavior and interactions after the quench. Also systems with multiple species (for e.g, two types of bosons) are realized in
low-dimensional cold atomic experiments 
\cite{thalhammer2008,ds1,ds2} and therefore, investigating the possible mapping to multiple coupled copies of chiral differential equations (for e.g, coupled KdV equations) could be timely and important. Understanding the interplay between chiral differential equations could
shine light on the complex dynamics of multi-species systems. 

Clearly, we plan to compare our theory with experiments very soon and this will provide input to improve our modeling. 
A more ambitious future direction would be to study quenches of excited states of quantum systems in regimes where they do not admit a hydrodynamic description and to discuss the behaviour of quantum states that in the hydrodynamical regime behaves like a soliton.

\section{Acknowledgements}
We acknowledge discussions with A. G. Abanov, A. Gromov, 
A. Polkovnikov, A. Polychronakos, D. Schneble, S. Sotiriadis and G. Takacs.
FF was supported by a Marie Curie International Outgoing 
Fellowship within the 7th European Community Framework Programme 
(FP7/2007-2013) under the grant PIOF-PHY-276093. FF acknowledges support from the H2020 Twinning project No. 692194, ``RBIT-WINNING''. FF and AT acknowledge support 
from the European Project Matterwave. 
MK gratefully acknowledges support from the Professional 
Staff Congress $-$ City University of New York award \# 68193-00 46. 
AT acknowledges support from the Italian PRIN ``Fenomeni quantistici collettivi: dai sistemi fortemente correlati ai simulatori quantistici''.
MK would also like to thank the hospitality of the International Centre for Theoretical Sciences of the Tata Institute of Fundamental Research, Bengaluru.

\appendix

\section{Universality of the KdV equation}
\label{app:KdVreduction}

The existence of solitons has been foremost 
an empirical observation \cite{newell}. 
It then took a while to realize that solitons appear as solutions 
of integrable differential equations. The reason for which 
solitonic waves can propagate in actual physical systems is 
that under general assumptions it is possible to isolate an 
integrable core in the dynamics, while the rest of the 
terms can often be neglected up to a given time scale. 
The typical appearance of this ``non-linear universality'' 
is the emergence of the KdV equation 
as the integrable core of local fluid for shallow waves \cite{newell}.

In \cite{kulkarni2012} it was reproduced the standard 
derivation of KdV in classical fluids to extend it to the hydrodynamic 
treatment of cold quantum systems. We review in this Appendix 
their approach, to set the notations that we used in the body of this work.

Our starting point are the continuity and Euler equations (\ref{cont}), 
which assume no dissipation. The results are not modified 
qualitatively, as long as dissipative effects enter linearly.
We introduce the following notations
\bea
A_0 &=& A(\rho_0) \; ,
\label{A0def}\\
\omega_0 &=& \omega(\rho_0) \; , \\
\xi_{\pm} &=& x\pm c t \; ,
\eea
with respect to the equilibrium density $\rho_0$.
We want to describe long wave excitations on top of the constant background density $\rho_0$. Thus we are looking for solutions in the form
\bea
\rho (x,t) &=& \rho_0 + \delta \rho_{\alpha}(\xi_\alpha,t) \; , \\
v(x,t) &=& \delta v_{\alpha}(\xi_\alpha,t)  \; ,
\eea
where $\alpha = \pm$ indicates the wave chirality and where we anticipated the fact that small waves will move with velocities close to the sound speed. 

The perturbative expansion is based on a counting scheme with formal counting parameter $\epsilon$. We introduce the expansion of velocity and density fields as
\bea
\delta \rho_{\pm}(\xi_\pm,t) &=& \epsilon^2 \rho^0_\pm (\epsilon\xi_\pm,\epsilon^3t) + \epsilon^4 \rho^1_\pm(\epsilon\xi_\pm,\epsilon^3t) +\ldots\\
\delta v_{\pm}(\xi_\pm,t) &=& \epsilon^2 v^0_\pm (\epsilon\xi_\pm,\epsilon^3t) + \epsilon^4 v^1_\pm (\epsilon\xi_\pm,\epsilon^3t) +\ldots
\eea
We substitute this ansatz into (\ref{cont},\ref{Euler}) and collect terms $\sim \epsilon^3$ and $\sim \epsilon^5$. At cubic order we get 
\be \label{sol1}
v^0_\pm = \mp \frac{\omega^\prime_0 }{c} \rho^0_\pm 
\ee
and the consistency 
\be
c^2 = \rho_0\omega^\prime_0 \; .
\label{KdVc}
\ee

At the next order, we can combine the two hydrodynamical equations to get
\be
\dot{u}_\pm \mp  \zeta u_\pm \partial_x u_\pm  \pm \alpha \p^3_x u_\pm = 0 \; ,
\label{KdV}
\ee
where we introduced
\bea
\rho^0_\pm & \equiv  & u_{\pm} \; , \\
\zeta & \equiv & \left(\frac{3\omega_0^{'}}{2c} + \frac{\omega^{\prime\prime}_0\rho_0}{2c}\right) 
=  \frac{c}{\rho_0} + \frac{\p c}{\p \rho_0} \; , 
\label{zetadef} \\
\alpha & \equiv & \frac{A_0}{4c} 
\label{alphadef} \; .
\eea
We also used the identity (\ref{KdVc}) so that 
\be
\frac{\p c}{\p\rho_0} = \frac{\omega^{\prime}_0}{2c} +\frac{\omega^{\prime\prime}_0\rho_0}{2c} = \frac{c}{2\rho_0} +\frac{\omega^{\prime\prime}_0\rho_0}{2c} \; .
\ee 
Shifting (\ref{KdV}) back from the reference frame moving with the sound velocity to the laboratory frame, we have (\ref{2KdV}).

In this derivation, we considered separately the two chiral sectors. A generic initial condition, however, will consist of both chiralities and, in principle, one should take into account the interaction between them. However, these effects can be neglected since, due to locality, the two sectors interact only when they are overlapping, but this happens for a short time, as they pass through each other with a relative velocity of approximately $2c$.

We have thus shown that the dominant non-linear contributions to the dynamics of shallow, long waves in a generic one-component hydrodynamic system (\ref{cont}) is given by (\ref{2KdV}), which is known as the integrable KdV equation.

\section{Generalities on KdV}
\label{app:KdV}

The non-linear term $\zeta$ in the KdV (\ref{2KdV},\ref{KdV}) pushes the different parts of $u(x,t)$ to move with different velocities, so that small perturbations over the background move close to the speed of sound, while the parts more distant from the asymptotic equilibrium $\rho_0$ move with higher $\delta V$. This term tends to generate a shock-like profile, as the tip of $u$ moves faster than the base. The dispersive, $\alpha$, term, instead, redistributes the kinetic energy within the $u$ profile and effectively lead to a broadening of the $u$ profile.

We can introduce two time scales capturing the effective strength of these two terms
\bea
\Omega_{\alpha} &=& \alpha W^{-3}\; , \\
\Omega_{\zeta} &=& \zeta U W^{-1} \; ,
\eea
where $W$ is the typical width of the disturbance over the background and $U$ is its typical size (note that $U$ has the unit of an inverse length, as it describes the height of a density bump).
There are three possible regimes
\be
\Omega_{\alpha} << \Omega_{\zeta}, \qquad \Omega_{\alpha} >> \Omega_{\zeta} \qquad {\mbox{and}} \qquad \Omega_{\zeta} \sim \Omega_{\alpha}.
\ee
Physically, they correspond to a regime of dominant non-linearity (true non-linear KdV dynamics), dominant dispersion with linear evolution (dispersive waves) and solitonic (equilibrium between non-linearity and dispersion). 
The dimensionless ratio 
\be
\frac{\Omega_{\zeta}}{\Omega_{\alpha}} = \frac{ \zeta }{ \alpha } U W^{2} 
\label{dimrat}
\ee
distinguish the non-linear ($\frac{\Omega_{\zeta}}{\Omega_{\alpha}} \gg 1$), the dispersive ($\frac{\Omega_{\zeta}}{\Omega_{\alpha}} \ll 1$), and the solitonic regime ($\frac{\Omega_{\zeta}}{\Omega_{\alpha}} \sim 1$).

The KdV supports true solitons which keep a perfect equilibrium between the non-linear and dispersive effect and thus propagate without changing their shape. The solitons can be both bright (density higher than background) or gray (a depletion in density). In fact, (\ref{KdV}) is invariant under the simultaneous reversal of the sign and chirality of $u_\pm \to - u_\mp$. 
The single (dark) soliton solutions has the form
\be
s(\xi_\pm,t) = - U \cosh^{-2} \left[ {\xi_\pm \mp \delta V t \over W} \right] \; ,
\label{soldef}
\ee
where $\delta V = c-V$ is the velocity of the soliton (in the sound velocity frame). The height and width of the depletion depend on the soliton velocity as
\bea
W & = & 2 \sqrt{\alpha \over \delta V} \; , 
\label{Wdef} \\
U & = & 3 {\delta V \over \zeta} \; .
\label{Udef}
\eea
The integrals of motions are
\bea
I_0 & = & n = \zeta \int u(\xi,t) d \xi = - 2 \zeta U W = - 12 \sqrt{\alpha \delta V} \; ,
\label{ndef} \\
I_1 & = & \zeta^2 \int {1 \over 2} u^2 (\xi,t) d \xi = {2 \over 3} \zeta^2 U^2 W = - \delta V \: n \; , 
\label{I1def}\\
I_2 & =& \zeta^2 \int \left[ \alpha u_\xi^2 (\xi,t) + {\zeta \over 3} u^3 (\xi,t) \right] d \xi 
= - {4 \over 15} \zeta^3 U^3 W 
= {6 \over 5} \delta V^2  \; n \; ,
\eea
where we rescaled everything by the right factors of $\zeta$, which sets the scale of the amplitude wave, see (\ref{2KdV}).

We note that the soliton velocity can be found as the ratio between its momentum $I_1$ and its ``mass'' $I_0$:
\be
\delta V = \zeta { \int {u^2 \over 2} de\xi \over \int u d \xi}  \; ,
\label{deltaVest}
\ee
as can be expected by averaging the nonlinear term in the KdV (\ref{KdV}), see also the comment after (\ref{Vofx}).

In section \ref{sec:quench} we considered the evolution of an initial condition which is functionally the same as the soliton (\ref{soldef}), but not necessarily with the correct parameters $U,W$ of a soliton.
The different parts of the KdV equation give:
\bea
u_x & = & {2 U \over W} {\sinh \left[ (x- V t)/ W \right] \over \cosh^3\left[ (x- V t)/ W \right] } \; , \\
\dot{u} & = & - V u_x \; , \\
u_{xxx} & = & {4 \over W^2} \left( 1 + {3 u \over U} \right) u_x \; .
\eea
We notice that for an inverse cosh square initial condition  every term in the KdV is proportional to its space derivative. Hence, we have
\be
 c u_x + \zeta u u_x - \alpha u_{xxx} =
 \left[ c   - {4 \alpha \over W^2} +  \left( \zeta - {12 \alpha \over U W^2} \right) u(x-Vt) \right] u_x \; .
\label{Vofx}
\ee
For a soliton, $U W^2= {12 \alpha \over \zeta}$ and the terms multiplying $u_x$ become a constant, which equals the soliton velocity. For generic $U$ and $W$, the terms within parenthesis give the velocity of each part of the profile. 

Note that the center of mass of a profile of the type (\ref{soldef}), that is the height at which the profile has equal area above and below it, is at a third of its total amplitude. In fact the velocity at $u = U/3$ from (\ref{Vofx}) is $c - {\zeta \over 3} U = V$, that is, the bulk velocity of the profile.

\section{The Non-Linear Schr\"odinger Equation and its KdV Reduction}
\label{app:NLSE}

The Non-Linear Schr\"odinger equation (NLSE) is characterized by 
the dynamical equation (\ref{NLS}).
The asymptotic modulus of the field 
$\lim_{x \to \pm \infty} |\Psi (x,t)|^2 = \rho_0$, which defines the (asymptotic) density of particle, helps in defining the dimensionless parameter
\be
\gamma \equiv \frac{m}{\hbar^2} \frac{g}{\rho_0} \; ,
\label{gammadef}
\ee
which controls the effective strength of the interaction. In the weakly interacting limit $\gamma \ll 1$, the NLSE captures the dynamics 
of the Lieb-Liniger model, describing an integrable system of one dimensional bosons with contact interaction (see the discussion in Section \ref{sec:hydro}).

From (\ref{NLSEparameters}) 
we can extract the phenomenological parameters of the KdV as
\bea
c &=& \sqrt{g \rho_0 \over m} 
= \frac{\hbar}{m}\rho_0\sqrt{\gamma} \; , 
\label{cNLS}\\
\zeta &=& {3 \over 2} \sqrt{ g \over m \rho_0} 
= \frac{\hbar}{m} \frac{3}{2} \sqrt{\gamma} 
= {3 \over 2} {c \over \rho_0} \; , 
\label{zetaNLS}\\
\alpha &=& {\hbar^2 \over 8 m^2} \sqrt{m \over g \rho_0} 
= \frac{\hbar}{m} \frac{1}{8\rho_0} \frac{1}{\sqrt{\gamma}}
\label{alphaNLS} \; .
\eea

The NLSE is also integrable and its single (dark) soliton solutions is 
\cite{stringari,novikov1984}
\be
\psi (x,t) = \sqrt{\rho_0} \left\{ {V \over c} 
- i \sqrt{ 1 - {V^2 \over c^2} }
\tanh \left[( x - V t) \sqrt{ {g m \over \hbar^2} \left(1 - {V^2 \over c^2} \right) \rho_0}   \right] \right\} \; .
\label{NLSESoliton}
\ee 

We have 
\be
\delta \rho (x,t) = |\psi (x,t)|^2 - \rho_0 = 
- \left( 1 - {V^2 \over c^2} \right) \: \rho_0 \cosh^{-2} \left[( x - V t) \sqrt{ {g m \over \hbar^2} \left(1 - {V^2 \over c^2} \right) \rho_0}   \right] \; .
\ee
We notice that the NLSE soliton has the same functional form as that of the KdV (\ref{soldef}), with width and height given by
\bea
W = {1 \over  \sqrt{ {g m \over \hbar^2} \left(1 - {V^2 \over c^2} \right) \rho_0}}
& \simeq & {1 \over \rho_0} \sqrt{ c \over 2 \gamma \delta V} + \ldots   \; ,
\label{NLSEW} \\
U =  \left( 1 - {V^2 \over c^2} \right) \rho_0  
& \simeq & 2 {\delta V \over c} \: \rho_0 + \ldots \; , 
\label{UNLS}
\eea
where we expanded for soliton velocities close to the speed of sound and found consistency with (\ref{Wdef}, \ref{Udef}). 

We conclude this Section observing that with a power-law non-linearity the NLSE reads
\be
i \hbar \partial_t \Psi (x,t) = \left\{ - {\hbar^2 \over 2 m} \partial_{xx} 
+ g \left| \psi (x,t) \right|^{2(\kappa-1)} \right\} \psi(x,t) \; .
\label{mNLS}
\ee 
If the NLSE is intended to mimic $2$-body interactions, this generalization considers $\kappa$-body contact interaction, i.e. a term ${g \over 2 \kappa} \: |\psi(x,t)|^{2 \kappa}$ in the Hamiltonian.
The hydrodynamic parameters are in this case
\bea
\omega(\rho) &=& \frac{g}{m}\rho^{\kappa -1} \; ,\\
A &=& \frac{\hbar^2}{2m^2} \; ,
\eea
for which we find that the corresponding KdV has 
\bea
c &=& \sqrt{(\kappa -1) g \rho_0^{\kappa -1} \over m}  \; , \\
\zeta &=& {\kappa + 1 \over 2} \sqrt{ (\kappa -1) g \rho_0^{\kappa -3} \over m } 
= {\kappa + 1 \over 2} {c \over \rho_0} \; , \\
\alpha &=& {\hbar^2 \over 8 \: c \: m^2}  \; .
\eea

\section{Calogero Model}
\label{app:Calogero}

The Calogero model \cite{Calogero-1969,Sutherland-1971} is defined by the Hamiltonian
\be
H = {1 \over 2 m} \sum_{j=1}^N p_j^2 + {\hbar^2 \over 2 m} 
\sum_{j \ne k} {\lambda^2 \over (x_j - x_k)^2} \; ,
\ee
with a dimensionless coupling constant $\lambda$. The hydrodynamic description of this model has parameters \cite{kulkarni2012}
\bea
\omega (\rho) & =& {\hbar^2 \lambda^2 \over m^2} \left[ {1 \over 2} (\pi \rho)^2 + \pi \rho_x^H \right] \; , \\
A (\rho) & = & {\hbar^2 \lambda^2 \over 2 m^2} \; ,
\eea
where the superscript $H$ stands for the Hilbert transform:
\be
f^H (x) = {1 \over \pi} \dashint {f(y) \over y - x} \: d y \; .
\ee

Due to the long-range nature of the Calogero interaction, its hydrodynamic description does not reduce to a KdV equation, but is given by the so-called 
{\it double Benjamin-Ono} \cite{abanov2009}, which, for small (chiral) profiles reduces to the usual Benjamin-Ono equation
\be
u_t \pm \left[ c \: u_x +  \zeta u u_x + \alpha_{BO} \left( u_{xx} \right)^H \right] = 0 \; ,
\label{BOeq}
\ee
with parameters
\bea
c & = & {\hbar  \: \pi \: \lambda \over m} \: \rho_0 \; , 
\label{BOcdef} \\
\zeta & = & 2 \: {\hbar \: \pi \: \lambda \over m} = {2 c \over \rho_0} \; ,
\label{BOzetadef} \\ 
\alpha_{BO} & = & {\hbar \: \lambda \over 2 m} = {c \over 2 \pi \rho_0} \; .
\label{BOalphadef}
\eea
The one-soliton density profile \cite{ap1994} for the rational Calogero is
\be
s_C= {\rho_0 \tau \over \left[ \pi \rho_0 (x - V t) \right]^2 + \tau^2} \; ,
\label{calsol}
\ee
where
\be
\tau = {c^2 \over V^2 - c^2}  \; .
\label{DoubleBOtau}
\ee
Note that for this model, the soliton is bright (positive density displacement) and hence its velocity $V >c$ is supersonic.
The height and width of (\ref{calsol}) are given by
\bea
U & = & {\rho_0 \over \tau} = \rho_0 \left( {V^2 \over c^2} - 1 \right) \; , \\
W & = & {\tau \over \pi \rho_0} = {1 \over \pi U} \; .
\eea

Let us again consider the evolution of an initial profile like (\ref{calsol}), but with generic parameters, such as
\be
u= {a \over \left[ \pi \rho_0 (x - V t) \right]^2 + b^2} \; .
\label{calsollike}
\ee
We have
\bea
u_x & = & - 2 (\pi \rho_0)^2 {a (x-Vt) \over  \left( \left[ \pi \rho_0 (x - V t) \right]^2 + b^2 \right)^2 }\; , \\
\dot{u} & = & - V u_x \; , \\
\left( u_{xx} \right)^H & = & \pi \rho_0 \left( {1 \over b} - {4 b \over a} \: u \right) u_x \; . 
\eea
Hence, the velocity of each part of the profile can be read from the BO equation as
\be
  c \: u_x + \zeta u u_x + \alpha_{BO} \left( u_{xx} \right)^H =
 \left[ c + \alpha_{BO} \:  {\pi \rho_0 \over b} + \left( \zeta - \alpha_{BO}{4 \pi \rho_0 b \over a} \right) u \right] \: u_x  \; .
\label{varVBO}
\ee

We see that for a soliton ($a = \rho_0 b$) the term proportional to $u$ vanishes and the soliton moves with velocity $c \left( 1 + {1 \over 2 b} \right)$, or
\be
  b = {1 \over 2} {c \over V -c} \; , 
  \label{bV}
\ee
which reproduces (\ref{DoubleBOtau}) only to leading order in $\delta V = V-c$.
Thus, we notice a mismatch between the parameters of the soliton supported by (\ref{BOeq}) and those of the true Calogero soliton (\ref{calsol},\ref{DoubleBOtau}), which is propagated by the double BO. In the limit of small $\delta V$ the two approach,
in the same way as the KdV is a valid approximation to a 
local hydrodynamics only for shallow waves.

Similarly to what we did for the KdV, we can find the velocity of the profile (\ref{calsollike}) from the conserved quantities, as in (\ref{deltaVest}). We have
\bea
\int u \:d x & =& {a \over \rho_0 b} \; , \\
\int {u^2 \over 2} \: d x & = & {a^2 \over 4 \rho_0 b^3} \; .
\eea
Hence, the average height of the profile (\ref{calsollike}) is
\be
\langle u \rangle = { \int {u^2 \over 2} \: d x \over \int u \: d x } =
{a \over 4 b^2} \; \,
\label{avcal}
\ee
and its bulk velocity is
\be
V_{\rm av} = c + \zeta  \langle u \rangle 
= c \left( 1 + { a \over 2 \rho_0 b} \right) \; .
\label{Vav}
\ee
Eq. (\ref{calsollike}) becomes a soliton of (\ref{BOeq}) for $a = \rho_0 b$ and in that case (\ref{Vav}) correctly reproduces (\ref{bV}).
The velocity (\ref{Vav}) coincides with the velocity at the center-of-mass, calculated from (\ref{varVBO}) at the height (\ref{avcal}), since at this height the dispersive effects are perfectly balanced and cancel out.

The center of mass velocity can be compared with the velocity at the peak by setting $u = a/ \tau^2$ in (\ref{varVBO})
\be
V^{\rm Peak} = c + \zeta {a \over b^2} - \alpha_{BO} {3 \pi \rho_0 \over b}
= c \left( 1 + {2 a \over \rho_0 b^2} -  {3 \over 2 b} \right) \; .
\label{BOVpeak}
\ee

\section{Harmonic Calogero Model}
\label{app:HarmCal}

The Calogero interaction remains integrable even when an external harmonic potential is applied, as in (\ref{harmCal}).
Most of what is valid for the rational case considered above remains valid, but the background density is now not constant and follows the famous semicircle 
law \cite{abanov2011}
\be
\rho_0 (x) = {\omega \over \pi \lambda} \sqrt{R^2 - x^2} \; , \qquad {\rm where} \qquad
R \equiv \sqrt{ 2 \lambda N \over \omega} \; .
\label{semicircle}
\ee
Thus, the speed of sound is not a constant, since it depends on the local density, and 
is given by
\bea
 c(x)= {\hbar \over m} \pi \lambda \rho_0 (x)= {\hbar \over m} \omega \sqrt{R^2 - x^2}.
\eea
We notice that the speed of sound at the center of the trap ($x=0$ ) is
\be
c = \sqrt{2 \lambda \omega N} \; ,
\ee
and it decreases as we move from the center.

The soliton solutions of the harmonic model were found recently \cite{abanov2011}. They can be thought of as ``Lorentzians'' like (\ref{calsollike}) that live on top of the background density (\ref{semicircle}):
\begin{eqnarray}
	\rho(x,t) & = & \rho_{0}(x)+\frac{1}{\pi}\frac{y_{1}(t)}{[x-x_{1}(t)]^{2}+y_{1}^{2}(t)}
 \label{eq:rsol}\\
	v(x,t) & = & -g\frac{y_{1}(t)}{[x-x_{1}(t)]^{2}+y_{1}^{2}(t)},
 \label{eq:vsol}
\end{eqnarray}
where 
\bea
z(t)=x_1(t)+i y_1(t)
\eea
is an external parameter that drives the soliton as it traces an ellipse in the complex plane (see also Ref. \onlinecite{abanov2011} for further details on the external parameter which we dub as dual variable)
\be
	z (t) = z(0) e^{i\omega t} + \frac{\sin \omega t}{\omega}\left[P(0)-i\omega X(0)\right] \; .
 \la{1solellipse}
\ee
Here $z (0)$ is the initial position of $z_{1}$ in the complex plane and $X=\sum_{j=1}^{N}x_{j}$, $P=\sum_{j=1}^{N}p_{j}$ are the center of mass and the total momentum of the system at $t=0$. Without loss of generality, we can take $z (0)=i b$ with $b>0$ as initial condition, which also gives $X=0$.
The equation of the ellipse in this case is
\be
	z(t)=ib \cos(\omega t)-b\Big(1+\frac{1}{\omega}\sum_{j}\frac{\lambda}{x_{j}^{2}+b^{2}}\Big)\sin(\omega t),
 \la{explicitellipse}
\ee
We see that the initial value $b$ uniquely characterizes the soliton. 
Combining Eqs. (\ref{eq:rsol}) and (\ref{DoubleBOtau}), the soliton velocity at the center is given by 
\bea
V=c\sqrt{1+\frac{1}{\pi b \rho_0(0)}}=c\sqrt{1+\frac{1}{b}\sqrt{\frac{\lambda}{2\omega N}}}
\label{vt1}
\eea

The soliton velocity changes as it moves away from the center (i.e, as a function of time). Since the soliton follows the external complex parameter $z(t)$, its velocity matches that of the real part of the dual variable $z(t)$: 
\bea
V(t)= \Big[1+\sum_{j}\frac{\lambda}{x_{j}^{2}+b^{2}} \Big]b \cos (\omega t)
\label{vt2}
\eea
By inspection of Eq. (\ref{vt1}) and Eq. (\ref{vt2}) we get,
\bea
V(t)= c\sqrt{1+\frac{1}{b}\sqrt{\frac{\lambda}{2\omega N}}} \cos (\omega t)
\label{vt3}
\eea
As one can notice from the above equation, the soliton velocity decreases as it moves away from the center. It is also worthwhile noticing that the soliton width, $y_1(t)$, also decreases. 
Therefore, we have a scenario where a soliton is moving slower, as it becomes thinner.  
In flat background, thinner solitons move faster and thus we see that the interesting interplay between the non-constant background and the soliton moving on top it contradicts the common wisdom valid in constant background.

\end{document}